\documentclass[aps,prl,reprint,amsmath,amssymb,superscriptaddress,showkeys,longbibliography]{revtex4-2}
\usepackage{bm,psfrag,setspace,bbold,braket}
\usepackage{mathrsfs}
\usepackage{amsthm}
\usepackage{hhline}
\usepackage[T1]{fontenc}
\usepackage[left]{lineno}
\usepackage[pdftex]{graphicx}
\usepackage{accents}
\usepackage{hyperref}
\usepackage[usenames,dvipsnames]{color}
\hypersetup{colorlinks=true,linkcolor=blue,citecolor=blue,urlcolor=blue,hypertexnames=true}
\selectfont

\begin{document}
\setlength{\parskip}{0pt}
\title{Deterministic photonic entanglement arising from non-Abelian quantum holonomy}
\author{Aniruddha Bhattacharya}
\email{Contact author: anirudb@umich.edu, he/him/his}
\affiliation{School of Physics, Georgia Institute of Technology, 837 State Street, Atlanta, Georgia 30332-0430}
\author{Chandra Raman}
\email{Contact author: chandra.raman@physics.gatech.edu, he/him/his}
\affiliation{School of Physics, Georgia Institute of Technology, 837 State Street, Atlanta, Georgia 30332-0430}
\date{\today}

\begin{abstract}

Realizing deterministic, high-fidelity entangling interactions—of the kind that can be utilized for efficient quantum information processing—between photons remains an elusive goal. Here, we address this long-standing issue by devising a protocol for creating and manipulating highly-entangled superpositions of well-controlled states of light by using an on-chip photonic system that has recently been shown to implement three-dimensional, non-Abelian quantum holonomy. Our calculations indicate that a subset of such entangled superpositions are maximally-entangled, “volume-law” states, and that the underlying entanglement can be distilled and purified for applications in quantum science. Crucially, we generalize this approach to demonstrate the potentiality of deterministically entangling two arbitrarily high, $N$-dimensional quantum systems, by formally establishing a deep connection between the matrix representations of the unitary quantum holonomy—within energy-degenerate subspaces in which the total excitation number is conserved—and the $\left(2j+1\right)$-dimensional irreducible representations of the rotation operator, where $j = \left(N-1\right)/2$ and $N \geq 2$. Specifically, our protocol deterministically entangles spatially localized modes that are not only distinguishable but are also individually accessible and amenable to state preparation and measurement, and therefore, we envisage that this entangling mechanism could be utilized for deterministic quantum information processing with light.

\end{abstract}

\maketitle

\emph{Introduction.}—Quantum entanglement is among the most mysterious aspects of quantum mechanics \cite{bell2004speakable, RevModPhys.81.865}, and has become an essential resource for practically all quantum technologies \cite{nielsen2010quantum}. Ever since its conceptualization \cite{PhysRev.47.777, schrodinger1935gegenwartige, Schrödinger_1935}, received wisdom had suggested that entanglement occurs in two ways:  Either a pair of particles shares a common history, for example, having originated from the same source \cite{PhysRevLett.28.938, PhysRevLett.47.460, PhysRevLett.49.1804}, or they directly interact with each other physically \cite{PhysRevLett.79.1}. Surprisingly, it was later shown that entanglement could be generated by a broader class of methods.  For example, two particles can become entangled without interacting  by making measurements on a third system, i.e., by entanglement swapping \cite{PhysRevLett.70.1895, PhysRevLett.71.4287, PhysRevA.57.822, PhysRevLett.80.3891,Bouwmeester2000}. Moreover, two indistinguishable photons, without any shared history, that are incident at the input beam-splitter of a two-mode Mach-Zehnder interferometer—a linear optical counterpart of the Young's double-slit experiment—become entangled by interference, and exit the output beam-splitter as a path-entangled, or spatial ``mode-entangled'' two-photon state. In particular, the phenomenon of entanglement due to interference within one beam-splitter is known as the Hong-Ou-Mandel (HOM) effect \cite{PhysRevLett.59.2044}. While the HOM method is appealing for its linearity and unitarity, no obvious scheme has been devised to extend the entanglement to encompass larger photon numbers without introducing additional spatial modes \cite{walther2004broglie, Mahrlein:15}; post-selected, non-unitary operations \cite{knill2001scheme, PhysRevLett.89.137901, PhysRevLett.89.037904, PhysRevLett.91.120402, o2003demonstration, mitchell2004super}; or manifestly non-linear gates \cite{PhysRevA.64.063814}.

\begin{figure}[!]
\includegraphics[width=\columnwidth]{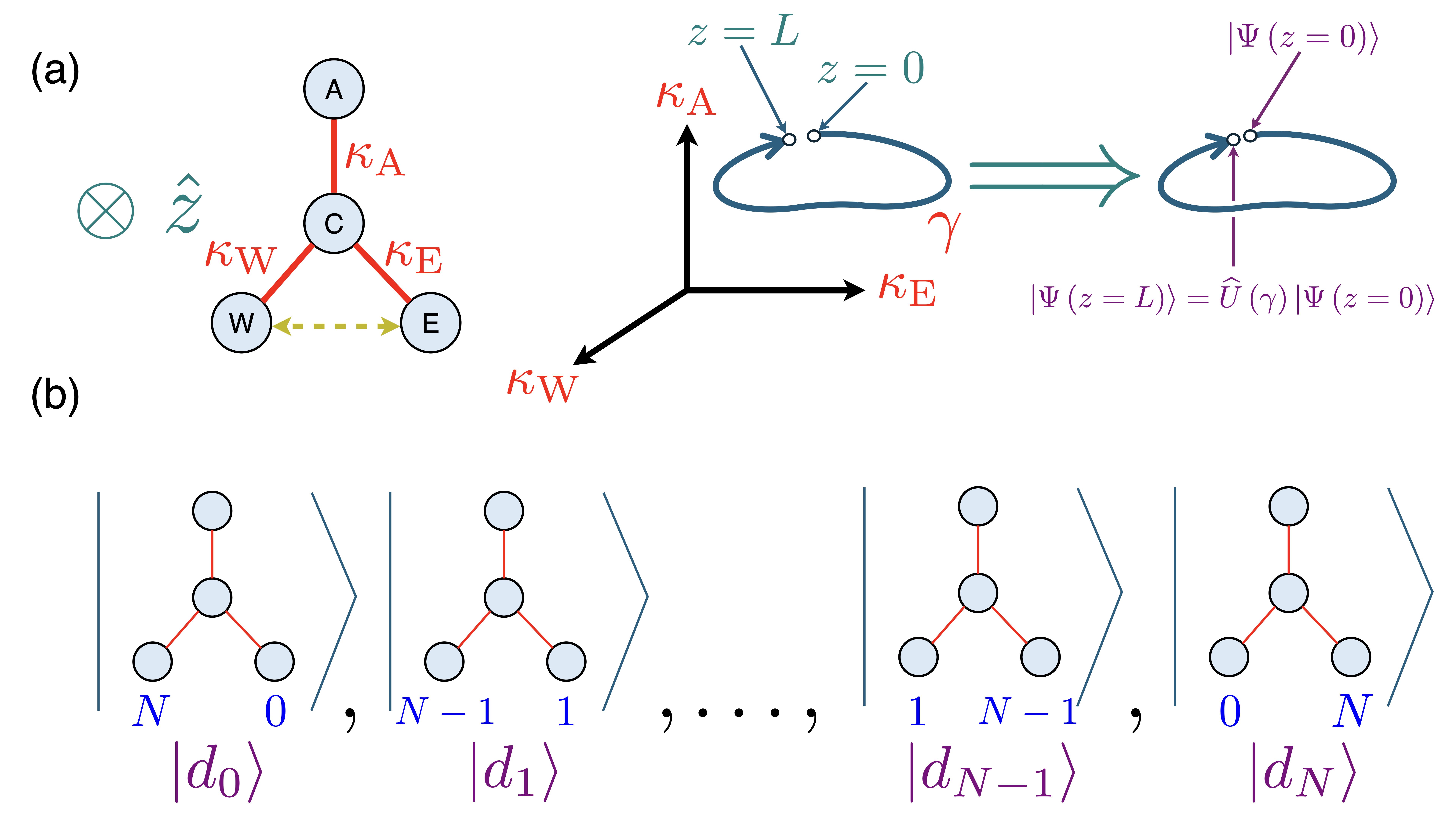}
\caption{\label{fig:f1} \textbf{High-dimensional, deterministic entanglement through holonomy.} (a) A system of four coupled optical waveguides: A, C, E, and W are shown in a transverse plane. The three inter-waveguide coupling coefficients, $\kappa_{\textrm{A},\textrm{E},\textrm{W}}$ vary with position $z$ along the length of the guide (in the propagation direction perpendicular to the plane of the page) in such a way as to define a closed path, $\gamma$ in $\kappa$ space, and thereby a unitary evolution, $\widehat{U}\left(\gamma\right)$. The relevant quantum subsystems are E and W at the input ($z=0$) and output ($z=L$) waveguide facets. (b) With $N$ input photons, this system realizes a non-Abelian quantum holonomy, while remaining adiabatically within a subspace of dark states, $\{\left\vert d_j \right\rangle\}$ of $\left(N+1\right)$ dimensions.  The basis states consists of $N$ photons distributed between two modes, i.e., $\left\vert d_j \right\rangle =\frac{1}{\sqrt{j!(N-j)!}}(\widehat{D}^{\dagger}_1)^{j}(\widehat{D}^{\dagger}_2)^{N-j}|\boldsymbol{0}\rangle$ for $j=0,\dotsc,N$. The two modes $\widehat{D}_1$ and $\widehat{D}_2$ coincide with $\textrm{E}$ and $\textrm{W}$, respectively at $z=0$ and $z=L$. In this work, we rigorously prove that entangled superpositions of such dark states are generated by the holonomy for arbitrary values of $N \geq 2$.}
\end{figure}

This Letter describes a theoretical approach to generating entangled states of arbitrary photon number without requiring either measurements or interactions. We transform an input, unentangled, two-mode Fock state of light by a single unitary process into maximally-entangled, linear superpositions of two-mode Fock states; notably, the entanglement is produced between distinguishable modes—as opposed to individual, identical particles—of light. This occurs as a consequence of the adiabatic evolution of highly-degenerate quantum systems along closed trajectories in a configuration space (see Fig. \ref{fig:f1}). Fundamentally, the combination of conservation of total photon number \cite{dalton2017quantum} and superposition of the relevant probability amplitudes \cite{RevModPhys.71.S288} through spatial, adiabatic propagation deterministically entangles the incident photons. It can, therefore, be viewed as a more generalized and a multi-photon analogue of the HOM principle; however, we emphasize that our entangling mechanism is not only fundamentally different but also creates a richer set of entangled states over and above the usual, highly-entangled, $N00N$ states.

Discovering and realizing scalable and controllable entangling protocols is key to developing useful quantum information processing platforms \cite{nielsen2010quantum}. Quantum computing with Rydberg atoms \cite{browaeys2020many}, for example, has evolved from the initial theoretical proposals of Rydberg blockade \cite{PhysRevLett.85.2208,PhysRevLett.87.037901} to utilization of entanglement for programmable, reconfigurable, many-qubit quantum processors \cite{bluvstein2022quantum, evered2023high, bluvstein2024logical}. On the other hand, the field of quantum information processing with linear optical elements alone has hardly progressed, despite the fact that photons were initially touted as the most promising candidate for quantum information processing \cite{PhysRevLett.62.2124} by virtue of their intrinsic—and thereby, directly and readily accessible—quantum-mechanical nature. The paradigmatic approach to quantum computing with linear optics suffers from being non-deterministic; furthermore, such approaches additionally require ancilla qubits and feedback from photo-detector outputs—that is, post-selection—to generate an effective non-linearity \cite{knill2001scheme, PhysRevA.68.064303, RevModPhys.79.135}. We note that the absence of deterministic linear-optical quantum computing schemes and the problem of deterministically preparing highly-entangled, multi-photon states are inextricably linked. Analogous approaches—such as, boson sampling \cite{aaronson2011computational,tillmann2013experimental,spring2013boson,PhysRevLett.119.170501} and resource-efficient linear-optical quantum computation utilizing cluster states \cite{PhysRevLett.86.910,  PhysRevLett.93.040503, PhysRevLett.95.010501}—have not yet succeeded in realizing truly universal, general-purpose, and programmable quantum machines in on-chip, photonic platforms. The usual approach to solving this problem has been to generate effective photon-photon interactions via strongly interacting atoms \cite{PhysRevLett.102.203902, peyronel2012quantum}; however, this method has not yet been successfully implemented either in semiconductor cavity QED systems or in on-chip, integrated photonics settings.

\emph{Our proposal.}—In this Letter, we do away with the above-mentioned requirement for single-photon non-linearity.  Instead, we utilize a physical mechanism based on non-Abelian, quantum holonomy to entangle photons.  Holonomy has recently been implemented in a system of coupled photonic waveguides on a fused silica chip \cite{neef2023three}.  Remarkably, we show that such systems of four coupled waveguides are unusually rich, and have previously unexplored potential for creating high-dimensional, photonic entanglement \cite{erhard2020advances, kysela2020path}. The deep connection between differential geometry—that is, holonomy—and adiabatic quantum systems—excited with multiple photons and operated within constant excitation number subspaces—is key to harnessing this potential.

Figure \ref{fig:f1} shows a schematic overview of this system. While earlier work has focused on demonstrating the holonomy in three dimensions with two input photons \cite{neef2023three}, here, we show a powerful result on entanglement by deriving an analytical expression for the matrix representation of the unitary holonomy, $\textbf{U}\left(N\right)$ for any dimension $N \geq 2$, provided that $\left(N-1\right)$ photons are introduced to the holonomic device. We accomplish this result by uncovering a powerful link between a four-waveguide system being operated in an $N$-dimensional, energy-degenerate subspace of ``dark'' states—for which there are no excitations in the central waveguide—and a high-dimensional pseudo-angular momentum vector, $\widehat{J}$ generating rotations in a subspace with fixed quantum number $j= \left(N-1\right)/2$. The holonomy matrix can then be compactly written as a rotation matrix whose action on input, two-mode, product states results in entangled superpositions of these states. To the best of our knowledge, this connection has not been established before, even though the importance of non-Abelian gauge fields to the adiabatic development of energy-degenerate quantum systems was made around 40 years ago by Wilczek and Zee \cite{PhysRevLett.52.2111}. Our work is made possible, in part, by the recent theoretical reformulation of photonic quantum holonomies in the Heisenberg picture \cite{PhysRevResearch.1.033117, PhysRevA.101.062314, neef2023three}.  Throughout this Letter, we will retain the principal terminology—such as, east, west, central, and auxiliary waveguides—that has been introduced in Ref. \cite{neef2023three}.

Notably, we demonstrate that this adiabatic, holonomic procedure can be used to deterministically entangle two photonic, quantum subsystems—each of which is defined in $N$-dimensional Hilbert spaces, where $N \geq 2$—and therefore, prepare maximally-entangled states of two $\textrm{qu}N\textrm{its}$ \cite{PhysRevLett.85.4418}. For the case of $N=3$, for example, we show that we can create maximally-entangled, bipartite states of qutrits—having $\textrm{log}_2 3$ ``e-bits'' of entanglement entropy—by tuning the value of the accumulated non-Abelian phase to $\approx 0.48 \; \textrm{rad},$ where the value of the phase is controllable, and is determined by the geometric dimensions and inter-waveguide coupling coefficients within the device. Our proposal is in concurrence with previous predictions that suggest that the underlying, enriched non-Abelian symmetry of certain quantum-mechanical systems can lead to a growth of the entanglement entropy \cite{PhysRevB.107.045102}.

\emph{The entangling mechanism.}—The spatial variation in the inter-waveguide couplings is engineered so as to synthesize a holonomy \cite{berry1984quantal, PhysRevLett.51.2167, PhysRevLett.52.2111} that belong to the symmetry group $\textrm{U}\left(N\right),$ where $N$ is determined by the excitation-level of the input, two-mode, Fock states. Under adiabatic propagation, any initial state that belongs to the $N$-fold energy degenerate subspace of dark states, and is introduced at the input facet of the waveguide transforms in accordance with the following unitary matrix that describes the $\textrm{U}\left(N\right)$ holonomy:
\begin{equation}  
\boldsymbol{U}\left(\gamma\right) = \textrm{exp}\left[-\phi\left(\gamma\right)\left(\boldsymbol{D}^{\dagger}_1\boldsymbol{D}_2-\boldsymbol{D}_1\boldsymbol{D}^{\dagger}_2\right)\right],
\label{eq:U_Matrix0}
\end{equation} 
where $\boldsymbol{D}^{\dagger}_1$ and $\boldsymbol{D}^{\dagger}_2$ are the matrix representations of the bosonic creation operators for the two-fold energy degenerate dark modes, and 
\begin{equation}  
\phi\left(\gamma\right)=\oint_{\gamma}\sin\varphi d\theta
\label{eq:Hol_Phase0}
\end{equation}
is the accumulated, non-Abelian, holonomic phase and is determined by the loop $\gamma$—in the curved parameter space—which is described by the angular parameters, $\theta$ and $\varphi$, each of which are, in turn, functions of the inter-waveguide coupling coefficients \cite{supp}. This unitary matrix is written in the following ordered basis—within a subspace having a fixed, total number of photons, $\left(N-1\right)$—of dark states: $\{\left|N, 0\right>; \; \left|N-1, 1\right>; \; \dotsc \; ;\left|1, N-1\right>; \; \left|0, N\right>\}.$ For these basis kets, the left and the right entries indicate the photon occupation numbers of the dark modes corresponding to the operators $\widehat{D}^{\dagger}_1$ and $\widehat{D}^{\dagger}_2$, respectively, as well as, the occupation numbers within the east and west waveguides, at the input and output facets, respectively. Notice that we have distinguished an operator from its matrix representation by bold-facing the symbol for the latter. The above input and output states at the waveguide facets are also energy eigenstates. Equation \ref{eq:U_Matrix0} is a direct consequence of the following theorem, whose derivation constitutes one of the central results of this Letter:

\textbf{The Representation of non-Abelian Holonomies Theorem.} The $N \times N$ real, orthonormal matrix representation of the quantum holonomy—where the dimension of the holonomy, $N$ is an integer strictly greater than 1—in the basis of the energy-degenerate dark states is equivalent to the $j = \left(N-1\right)/2$—or, the $\left(2j+1\right)$-dimensional—irreducible representation of the rotation operator, $\widehat{\mathscr{D}}\left(\alpha, \beta, \gamma\right) = \widehat{\mathscr{D}}_z\left(\alpha\right)\widehat{\mathscr{D}}_y\left(\beta\right)\widehat{\mathscr{D}}_z\left( \gamma\right)$, where $\alpha$, $\beta$, and $\gamma$ are the three Euler angles such that $\alpha = \gamma = 0$, and $j$ is the angular momentum quantum number.

The above theorem suggests that the unitary holonomy—within the dark state subspace—is equivalent to the rotation generated by a high-dimensional pseudo-angular momentum vector, $\widehat{J}$ within subspaces having fixed $j$. Such rotations are capable of generating entanglement without nonlinearities. We also show that $\beta=2\phi\left(\gamma\right)$ \cite{supp}. Notice that a rotation of a real angular momentum vector—such as the spin angular momentum—is incapable of producing such entangling interactions in the absence of nonlinearities. For example, spin squeezing—which is a form of metrologically-useful entanglement—is known to occur only in strongly interacting spin ensembles and requires Heisenberg interactions \cite{PhysRevLett.86.5870, leibfried2004toward, lee2024observation}. We emphasize that the phenomenon described in this paper is distinct from either single-qubit or global spin rotations. A detailed proof of the above theorem, along with an exploration of all of its consequences, is provided in the Supplemental Material \cite{supp}.

A unitary change in the basis of dark states—such as $\left|D_k\right> \mapsto \boldsymbol{S}\left|D_k\right>$, where $\boldsymbol{S}$ is an $N \times N$ unitary matrix—transforms $\boldsymbol{U}\left(\gamma\right)$ in a gauge-covariant manner to $\boldsymbol{S}\boldsymbol{U}\left(\gamma\right)\boldsymbol{S}^{-1}$. Therefore, although $\boldsymbol{U}\left(\gamma\right)$ is not manifestly gauge-invariant unlike in the Abelian case, all measured values remain invariant under such a gauge transformation. The symmetry of the star graph of four optical waveguides (see Fig. \ref{fig:f1}) that produces the enormous degeneracy of the dark states at each spatial coordinate along the length of the holonomic device is an example of a local, or a gauge symmetry. Correspondingly, there exists a connection, or a gauge field that effects the parallel transport of a state vector lying within this dark subspace. The definition of $\boldsymbol{U}\left(\gamma\right)$ in terms of this connection is the starting point of our derivation in the Supplemental Material \cite{supp}.

For the remainder of this Letter, we will demonstrate these effects for the case of $N = 3$, which requires input states of two photons. The case of $N = 4$ and its utility in constructing universal, entangling gates will be explored in a future publication. According to Eq. (\ref{eq:U_Matrix0}), the matrix that describes U(3) holonomy is:
\begin{widetext}
\begin{equation}
    \widehat{U}\left(\gamma\right) = \left[
        \begin{matrix}
             \cos^2\left(\phi\right) & 
             -\sqrt{2}\sin\left(\phi\right)\cos\left(\phi\right) & \sin^2\left(\phi\right)    \\
             \sqrt{2}\sin\left(\phi\right)\cos\left(\phi\right) & \cos\left(2\phi\right) & 
             -\sqrt{2}\sin\left(\phi\right)\cos\left(\phi\right)      \\
             \sin^2\left(\phi\right) & 
             \sqrt{2}\sin\left(\phi\right) \cos\left(\phi\right) & \cos^2\left(\phi\right)      \\
             \end{matrix}
    \right],
    \label{eq:U_Matrix}
\end{equation}
\end{widetext}
where $\phi \equiv \phi\left(\gamma\right)$ is the non-Abelian, holonomic phase.

\begin{figure*}
\includegraphics[width=\textwidth]{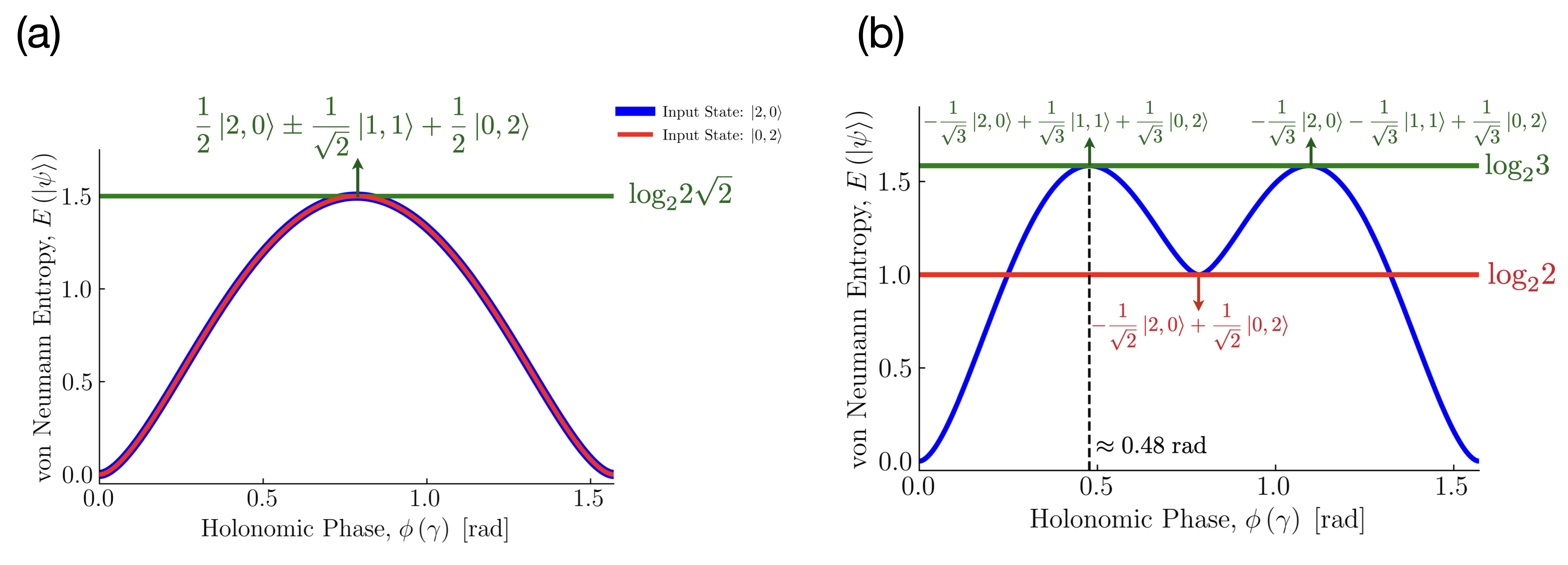}
\caption{\label{fig:f2} \textbf{Creating highly- and maximally-entangled photonic superpositions.} (a) The von Neumann entropy of entanglement of the output, pure state, $E\left(\left|\psi\right>\right)$ as a function of the non-Abelian phase, $\phi\left(\gamma\right),$ which is accumulated due to adiabatic propagation of the photons through the entire holonomic chip. The solid blue- and red-colored curves correspond to the situations in which the product states $\left|2,0\right>$ and $\left|0,2\right>$ are introduced at the input. The formation of the entangled, HOM-like states is shown by the green arrow. The solid horizontal line indicates the maximum possible value of entanglement entropy for a subsystem having a Hilbert-space of $2\sqrt{2}$ dimensions. (b) The same plot, as in (a), for the situation in which $\left|1,1\right>$ is introduced at the input. The vertical dashed line indicates the minimum value of non-Abelian phase that is required to create the maximally-entangled state. The formation of a few, well-known entangled superpositions are indicated; in particular, the ones in green are maximally-entangled. The solid horizontal lines indicate the maximum possible values of entanglement entropy for subsystem dimensions of $2$ and $3,$ respectively.}
\end{figure*}

\emph{The holonomic chip as an entangler.}—Suppose that we send in one of the basis states—that is, the two-photon dark states—which are factorizable, two-particle, product states, such as the $\left|2, 0\right>$ state. Consequently, according to Eq. (\ref{eq:U_Matrix}), the output state will be:
\begin{widetext}
\begin{equation}
    \left|\Psi\right>_{\textrm{out}} = \cos^2\left(\phi\right)\left|2, 0\right> + \sqrt{2}\sin\left(\phi\right)\cos\left(\phi\right)\left|1, 1\right> + \sin^2\left(\phi\right)\left|0, 2\right>,
    \label{eq:output_state}
\end{equation}
\end{widetext}
and by imparting the appropriate values of $\phi\left(\gamma\right)$, we can create the following two highly-entangled, HOM-like states:
\begin{subequations}\label{eq:ME_Output_States}
\begin{align}
    \left|\Psi\right>_{\textrm{out}}^{\phi = \pi/4} &= \frac{1}{2}\left|2, 0\right> + \frac{1}{\sqrt{2}}\left|1, 1\right> + \frac{1}{2}\left|0, 2\right>\label{eq:ME_Output_State1},\\
    \left|\Psi\right>_{\textrm{out}}^{\phi = 3\pi/4} &= \frac{1}{2}\left|2, 0\right> - \frac{1}{\sqrt{2}}\left|1, 1\right> + \frac{1}{2}\left|0, 2\right>\label{eq:ME_Output_State2}.
    \end{align}
\end{subequations}

For any bipartite system in a pure state, Bennett \emph{et al}. \cite{PhysRevA.53.2046} first showed that it is reasonable to quantify the above-mentioned degree of entanglement of the system as the von Neumann entropy of either of its two parts—that is, if $\left|\Psi\right>$ is the state of the whole system, then the entanglement can be defined as: $E\left(\left|\Psi\right>\right) = -\textrm{Tr}\left(\rho\textrm{log}_2\rho\right),$ where the density matrix, $\rho$ is the partial trace of $\left|\Psi\right>\left<\Psi\right|$ over either of the two subsystems. We will follow this definition throughout this Letter for pure states. In accordance with this definition, for example, separable product states have zero-valued entropies, whereas “maximally-entangled” states have an entropy of $\textrm{log}_2 \; N_{\textrm{H}}$ for subsystem dimensions of $N_{\textrm{H}}.$ More broadly, Fig. \ref{fig:f2} (a) shows the behavior of the entropy of entanglement as a function of the accumulated non-Abelian phase for the input state: $\left|2, 0\right>$ and its symmetric counterpart: $\left|0, 2\right>.$

\emph{Constructing maximally-entangled states.}—We now explore the situation in which the $\left|1,1\right>$ basis state is introduced at the input. Consequently, the output state will be:
\begin{widetext}
\begin{equation}                
 \left|\Psi\right>_{\textrm{out}} = -\sqrt{2}\sin\left(\phi\right)\cos\left(\phi\right)\left|2, 0\right> + \cos\left(2\phi\right)\left|1, 1\right> + \sqrt{2}\sin\left(\phi\right)\cos\left(\phi\right)\left|0, 2\right>\label{eq:11_state_output}.
\end{equation}
\end{widetext}
To maximize the degree of entanglement of this output state, we could equalize all of the magnitudes of the above expansion coefficients by requiring the overall non-Abelian phase to be:
\begin{equation}                
\phi\Bigg|_{\textrm{ME}} = \frac{1}{2}\tan^{-1}\sqrt{2} \approx 0.477 \; \textrm{rad}\label{eq:NAB_phase_ME},
\end{equation}
and therefore, the state that is produced at the output is:
\begin{equation}                
\left|\Psi\right>_{\textrm{out}}^{\textrm{ME}} = -\frac{1}{\sqrt{3}}\left|2, 0\right> + \frac{1}{\sqrt{3}}\left|1, 1\right> + \frac{1}{\sqrt{3}}\left|0, 2\right>\label{eq:11_state_output_ME}.
\end{equation}
The von Neumann entropy of entanglement of the above state is $\textrm{log}_2 3$ shannons; said differently, $\left|\Psi\right>_{\textrm{out}}^{\textrm{ME}}$ has $E\left(\left|\Psi\right>_{\textrm{out}}^{\textrm{ME}}\right) = \textrm{log}_2 3 \approx 1.58$ ``e-bits'' of entanglement. Notice that the value of $\phi\Bigg|_{\textrm{ME}} \approx 0.48 \; \textrm{rad}$ is within the realm of experimental possibility, as values of $\phi$ up to $\approx 0.33 \; \textrm{rad}$ have already been demonstrated. Moreover, a strategy for cumulatively enhancing this value of $\phi$, by sequentially cascading identical holonomic devices, is discussed in the Supplemental Material \cite{supp}. This enhancement in entanglement entropy—via adiabatic propagation through the holonomic chip—is reminiscent of how quantum adiabatic algorithms are used to prepare entangled states with high-fidelity \cite{RevModPhys.90.015002}.

The state described by Eq. (\ref{eq:11_state_output_ME}) is an entangled quantum state of $M = 2$ subsystems. Each of our subsystems has a dimension, $d = 3$—that is, the local, subsystem Hilbert space is isomorphic to $\mathbb{C}^3$—and consequently, the total, global Hilbert space of our bipartite system is $\mathcal{H}_{\textrm{tot}} = \left(\mathbb{C}^3\right)^{\otimes 2}.$ The closest analogue of this bipartite entangled state of two qutrits is the following tripartite Greenberger–Horne–Zeilinger state (GHZ state) of qutrits that also has an entanglement entropy of $\textrm{log}_2 3:$
\begin{widetext}
\begin{equation}                
 \left|\Psi\right>_{\textrm{GHZ}}^{M=d=3} = \frac{1}{\sqrt{3}}\left(\left|0\right> \otimes \left|0\right> \otimes \left|0\right> \right) + \frac{1}{\sqrt{3}}\left(\left|1\right> \otimes \left|1\right> \otimes \left|1\right> \right) + \frac{1}{\sqrt{3}}\left(\left|2\right> \otimes \left|2\right> \otimes \left|2\right> \right)\label{eq:GHZ_qutrits_3}.
\end{equation}
\end{widetext}
Our output state [Eq. (\ref{eq:11_state_output_ME})] saturates the maximum possible entropy of entanglement by virtue of being an equally distributed superposition of all possible product states in the two-photon subspace. Substituting $\phi = \pi/4$ in Eq. (\ref{eq:11_state_output}) recovers the conventional HOM state (see red arrow in Fig. \ref{fig:f2}) thereby supporting our interpretation of this holonomic device as a generalized version of the HOM interferometer. Figure \ref{fig:f2} (b) shows the generalized behavior of the entropy of entanglement as a function of the accumulated non-Abelian phase for the input state: $\left|1, 1\right>$; the maximally-entangled states, in particular, are indicated in green. Notice that all of the linear superpositions that are shown to be entangled—see, for example, Eqs. (\ref{eq:ME_Output_State1}), (\ref{eq:ME_Output_State2}), and (\ref{eq:11_state_output_ME})—also obey the global particle number super-selection rule \cite{dalton2017quantum, PhysRevA.67.013609, PhysRevA.68.042329}. An independent verification of this formation of bipartite, spatial entanglement through calculations of the second-order R\'enyi entropy as well as effects of quantum errors on such entangled states are described in the Supplemental Material \cite{supp} (see also references \cite{PhysRevLett.52.2111,neef2023three,PhysRevLett.51.2167,merzbacher1998quantum,sakurai2020modern,schwinger1952,RevModPhys.81.865, PhysRevA.54.1838, PhysRevLett.95.240407, horodecki1996quantum, PhysRevLett.23.880, islam2015measuring, brydges2019probing, wittig2005landau, PhysRevA.65.032314, PRXQuantum.2.030347} therein).

\emph{Outlook and conclusions.}—We envision that this holonomic entanglement can be used to realize universal, deterministic quantum processors. Specifically, in our approach, we will use the overall, system unitary transformation matrix—that is, the holonomy—as the unitary gate transformation matrix, and by tuning the non-Abelian phase, we can have this holonomy matrix realize—that is, be identical to, up to a global phase—those specific matrices that describe universal, entangling gates. Instead of realizing a qubit by having one photon in either of two optical modes—as is conventionally done in linear-optical quantum computation \cite{PhysRevA.52.3489, knill2001scheme, PhysRevLett.91.037903}—we will require a system of bi-partite qutrits, and eventually, higher-dimensional $\textrm{qu}N\textrm{its}$, where $N\geq4$. Furthermore, the holonomy confers added protection to these entangled states against the environment, since non-Abelian geometrical phases are less susceptible to dephasing-induced decoherence than their dynamical counterparts. A similar quantum architecture has been proposed \cite{PhysRevLett.129.160501}—for the quantum simulation of non-Abelian gauge theories—that uses optical tweezer arrays in which Rydberg atoms encode $\textrm{qu}d\textrm{its}$, e.g., $d=8$, and single- and two-qudit gates are performed holonomically \cite{zanardi1999holonomic, sjoqvist2012non, PhysRevLett.109.170501, PhysRevLett.110.190501, PhysRevA.97.042336, PhysRevA.103.052605}.

Our approach effectively brings the key, enabling ideas from the proposals for topological quantum computation \cite{RevModPhys.80.1083}—such as operating within highly energy-degenerate, decoherence-free subspaces—onto the realm of optical quantum computation. One future avenue of research, inspired by recent work on deterministically preparing non-Abelian, topologically ordered states \cite{PRXQuantum.4.020339, PhysRevLett.131.060405, PhysRevLett.131.200201, PhysRevX.14.021040, iqbal2024non}, might entail combining unitary evolution, quantum measurements, and feed-forward to produce long-range, multi-partite entangled states of photons, whose preparation is beyond the realm of scalable, unitary processes.

In summary, we have shown that an on-chip photonic device can effect entangling operations upon Fock states of photons by utilizing $\textrm{U}\left(N\right)$ non-Abelian, holonomic symmetry groups. We envisage that such entangling interactions between photons can be efficiently used to generate the conditional quantum dynamics that are needed for realizing universal, entangling quantum logic gates. The adiabatic propagation through the photonic chip is equivalent to the application of the ``operational pulse”—for example, the Rabi $\pi$ pulse in an all-optical quantum gate in semiconductor quantum dots \cite{li2003all}—that transforms a separable state into an entangled state. We conclude this Letter by emphasizing that our results indicate that non-Abelian quantum holonomic systems—printed on fused silica-based platforms—could become a robust, and possibly, integrable platform for quantum information processing with number states of light.

\emph{Note added.}—While we were preparing this manuscript, we became aware of a recently posted arXiv preprint \cite{neef2024non} that reports on the experimental realization of single-qubit quantum gates using non-Abelian, U(2) holonomies in this platform.

\begin{acknowledgments}
\emph{Acknowledgments.}—It is a pleasure to thank Sara Sloman for many stimulating discussions, during the initial stages of this work. This research was supported by the U.S. National Science Foundation (Award Number: 2011478). 
\end{acknowledgments}

\emph{Author Contributions.}—A.B. and C.R. conceptualized this work. A.B. carried out all the calculations and wrote the manuscript with feedback from C.R.

%

\section{End Matter}

\emph{Accessing higher-dimensional holonomies.}—The Heisenberg-based approach to realization of photonic quantum holonomies \cite{PhysRevResearch.1.033117, PhysRevA.101.062314, neef2023three} shows that the higher-dimensional holonomies, $\textrm{U}\left(N\right)$ [Eq. (\ref{eq:U_Matrix0})] are simply and directly instantiated by the following input, two-mode Fock states:
\begin{equation}
    \begin{split}
    \left|N_{\textrm{E}}, N_{\textrm{W}}\right\rangle &= 
    \left|N_1, N_2\right\rangle\Bigg|_{\substack{z = 0;\\ z = L}}\\ 
     &= \frac{1}{\sqrt{N_{\textrm{E}}!\,N_{\textrm{W}}!}}\left[\left(\widehat{D}^\dagger_1\right)^{N_{\textrm{E}}}  \left(\widehat{D}^\dagger_2\right)^{N_{\textrm{W}}}\right]\Bigg|_{\substack{z = 0;\\ z = L}} \left|\boldsymbol 0 \right\rangle, \\
    \end{split}
    \label{eq:N_Dim}
\end{equation}
where $N_{\textrm{E}} + N_{\textrm{W}} = N-1$ is the total number of photons. Equivalently, climbing up the symmetry group ladder, $\textrm{U}\left(N\right)$ is analogous to distributing $N-1$ identical bosons—that corresponds to the total number of input photons—in 2 distinguishable states, which gives rise to $\binom{N}{N-1} = N$ unique configurations. By suitably choosing a value of $\phi\left(\gamma\right)$, $N$ such Fock states can be linearly superimposed to produce maximal entanglement.

\begin{figure}[t]
\includegraphics[width=\columnwidth]{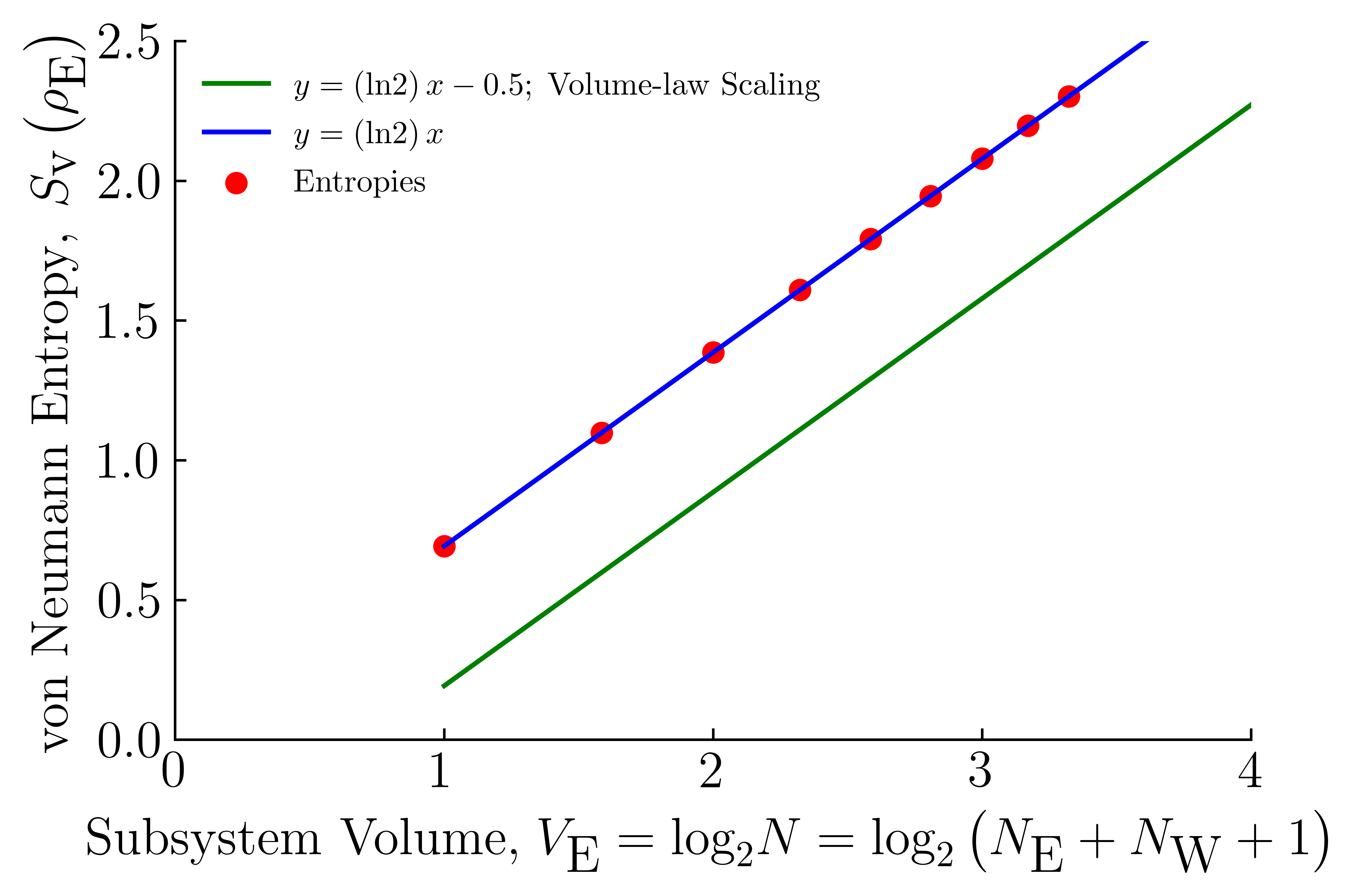}
\caption{\label{fig:f3} \textbf{Volume-law scaling of maximally-entangled states with holonomic dimension.} The variation in the von Neumann entropy of entanglement, $S_{\textrm{V}}\left(\rho_{\textrm{E}}\right)$ of one of the quantum subsystems—for example, the east waveguide—with the subsystem volume, which is the effective number of qubits that can be encoded in each of the subsystems. The subsystem volume scales with the logarithm—to the base 2—of the holonomic dimension, $N,$ which is equal to $N_{\textrm{E}}+N_{\textrm{W}}+1,$ where $N_{\textrm{E}}$ and $N_{\textrm{W}}$ are the photon occupation numbers—for the basis states—of the east and west waveguides, respectively. The red solid dots indicate the maximally-entangled states that can be generated in each of the holonomic dimensions by, for example, equalizing the expansion coefficients of the entangled superpositions. The solid green line represents the ideal volume-law scaling behavior, that is, the Page curve, whereas the solid blue line denotes $S_{\textrm{V}}\left(\rho_{\textrm{E}}\right) \propto V_{\textrm{E}}.$ The vertical shift between these two lines is a possible consequence of low dimensionality of the Hilbert space.}
\end{figure}

Moreover, as shown in Fig. \ref{fig:f3}, the above bipartite, spatially maximally-entangled states also follow volume-law scaling \cite{PhysRevLett.71.1291}. Although we have shown that this procedure of creating maximally-entangled states is true for $N \leq 3$—the case of $N=2$ is trivial—we conjecture that this method of varying the holonomic phase affords sufficient exploration of the Hilbert space so as to achieve equalization of the entanglement coefficients for higher dimensions as well.

\emph{Inter-mode and inter-particle entanglement between photons.}—Fundamentally, entanglement is a property of composite quantum systems, and consequently, the degree of entanglement of a particular state of a quantum system depends on the specific choice one makes in decomposing such a system into its constituent subsystems \cite{dalton2017quantum,10.1093/acprof:oso/9780199215706.001.0001,PhysRevA.66.052323,PhysRevA.67.013609,cunha2007entanglement,RevModPhys.81.865,GUHNE20091}. For example, the ground energy eigenstate of the hydrogen atom is entangled if we regard the individual positions of the electron and proton as the subsystems, and the same state is unentangled if the subsystems are the center of mass of the entire atom and the relative position of the electron and proton \cite{cunha2007entanglement}.

To describe the entanglement for systems of identical bosons, one could choose either the modes or the individual particles as the individual quantum subsystems. Given that well-defined quantum subsystems should be distinguishable; physically accessible and amenable to measurements \cite{RevModPhys.81.299}; and exist as separate systems that can be prepared in quantum states for that subsystem alone—the widely accepted perspective within the community is that distinguishable modes, as opposed to labeled indistinguishable particles, should be considered as subsystems for the notion of entanglement to be physically meaningful and measurable \cite{dalton2017quantum,10.1093/acprof:oso/9780199215706.001.0001,PhysRevA.66.052323,PhysRevA.67.013609,cunha2007entanglement,BENATTI2010924,Benatti_2011,walther2004broglie,Dalton_2014}. This framework is in consonance with the second quantization approach in which the basis states are the tensor products of Fock states—that automatically satisfy the symmetrization principle—of single particle states, which are the modes. The overall system is, therefore, a collection of modes, and particles are associated with the occupancies of such modes \cite{dalton2017quantum}. A concrete example of a spatial ``mode-entangled,'' superposition of photons is the following $N00N$ state:
\begin{equation}  
\left|\Psi\right>_{N00N} = \frac{1}{\sqrt{2}}\left(\left|N\right>_{\textrm{a}1}\left|0\right>_{\textrm{b}1} + e^{iN\Delta\varphi}\left|0\right>_{\textrm{a}1}\left|N\right>_{\textrm{b}1}\right),
\label{eq:N00N_State}
\end{equation}
where $\textrm{a}1$ and $\textrm{b}1$ represent the two spatial modes, respectively, and $N\Delta\varphi$ is the relative phase modulation. While such states—which are also described as path-entangled in photon number—can be created deterministically for $N=2$ photons with the HOM effect \cite{PhysRevLett.59.2044}, only a ``non-local'' version of the same can be prepared deterministically for $N=4$ with a Mach-Zehnder interferometer-like arrangement having four spatial modes \cite{walther2004broglie}.

Motivated by the above considerations, in our approach, we have chosen the spatially localized modes corresponding to the east and west waveguides—at the input as well as output facets—as our quantum subsystems [see Fig. \ref{fig:f1}(a)]. Crucially, such modes are distinguishable, physically accessible, and concrete, and therefore, can be used for both input state preparation and measurement of output results. Our protocol deterministically entangles identical photons that occupy such spatially localized, distinguishable modes, and crucially, the utility of our entangling protocol stems from the fact that these modes can be used to encode distinguishable qubits for the purpose of quantum information processing. Consequently, the entanglement that is generated between our preferred choice of subsystems is useful, as opposed to entanglement between two arbitrary basis representations of the modes that might be mathematically reasonable, but not necessarily physically accessible or utilizable. All of our calculations of the various measures of entanglement—such as, the von Neumann entropy in Figs. \ref{fig:f2} and \ref{fig:f3} of the main text, the second-order R\'enyi entropy in Fig. S2 of the Supplemental Material, and the logarithmic negativity in Fig. S3 of the Supplemental Material—have been consistently carried out by choosing the above-described modes, which are the most physically meaningful choice of the two quantum subsystems for our holonomic device.

We conclude this section with a brief overview of major shortcomings of the concept of particle entanglement in the context of the present discussion. First, regarding mathematically labeled individual identical particles as subsystems \cite{peres2002quantum,sorensen2001many,PhysRevA.86.012337,PhysRevLett.112.150501} is more of a mathematical—as opposed to a physical—approach, as such subsystems do not satisfy the above criteria for well-defined quantum subsystems \cite{dalton2017quantum}. In fact, it has been argued that the properly symmetrized wavefunctions of two identical particles are always entangled if one considers the individual particles as subsystems—a concept often referred to as entanglement due to symmetrization \cite{peres2002quantum}. Such entanglement due to symmetrization is usually neither useful nor utilizable for practical applications. While at least one scheme exists to extract the entanglement due to symmetrization \cite{PhysRevLett.112.150501}, the final measurements are, as before, based on entanglement of modes—as subsystems consisting of individually labeled identical particles are not measurable—and consequently, the consensus within the community is that entanglement due to symmetrization does not exist as a ``directly observable basic feature in composite quantum systems'' \cite{dalton2017quantum}.

\end{document}


\setlength{\parskip}{0pt}
\title{\texorpdfstring{Supplemental Material accompanying: \\ \emph{Deterministic photonic entanglement arising from non-Abelian quantum holonomy}}{Supplemental Material accompanying: \emph{Deterministic photonic entanglement arising from non-Abelian quantum holonomy}}}

\author{Aniruddha Bhattacharya}
\email{Contact author: anirudb@umich.edu, he/him/his}
\affiliation{School of Physics, Georgia Institute of Technology, 837 State Street, Atlanta, Georgia 30332-0430}
\author{Chandra S. Raman}
\email{Contact author: chandra.raman@physics.gatech.edu, he/him/his}
\affiliation{School of Physics, Georgia Institute of Technology, 837 State Street, Atlanta, Georgia 30332-0430}
\date{\today}

\maketitle

\tableofcontents

\renewcommand{\thesection}{S\arabic{section}}
\renewcommand{\theequation}{S\arabic{equation}}
\renewcommand{\thefigure}{S\arabic{figure}}

\section{Derivation of a Generalized Expression for the Unitary Holonomy Matrix}\label{sec:Derive}

\subsection{Introduction: Preliminaries and Notation}\label{sec:Intro}
We commence this introductory subsection by discussing several, relevant preliminaries, and thereafter, motivating the derivation of a generalized expression for the matrix representation of the holonomy that is true for any positive integral value of the holonomic dimension, $N \geq 2$. \textbf{Here, we will outline some key steps that we will undertake to facilitate the derivation of such a matrix. Subsequently, we will summarize our central results in the form of a theorem in Sec. \ref{sec:angular_momentum}.} Since, as we will discover later in this section, this matrix is real and orthonormal in our preferred choice of basis states, we will use the terms ``unitary" and ``orthogonal" interchangeably to describe these holonomic matrices.

The unitary transformation operator \cite{PhysRevLett.52.2111, neef2023three} that describes the quantum holonomy \cite{PhysRevLett.51.2167} is obtained from the following path-ordered integral:
\begin{equation}                
\widehat{U}\left(\gamma\right) = \mathcal{P}\textrm{exp}\left(\oint_{\gamma}\sum_{\mu}\widehat{A}_{\mu}d\kappa_{\mu}\right)\label{eq:U_matrix},
\end{equation}
where we have used the same notation as in Ref. \cite{neef2023three}, namely, $\gamma$ is a loop in the space of Hamiltonian coupling coefficients, $\{\kappa_{\mu}\}$, and $\sum_{\mu}\widehat{A}_{\mu}d\kappa_{\mu}$ is the adiabatic connection—analogous to the ones in quantum field theory—that brings about the parallel transport of the quantum state vector along the closed trajectory defined by $\gamma$. The individual components of the gauge field, $\widehat{A}_{\mu}$ are the gauge potentials. Throughout this Supplemental Material (SM)—as we have done in the main text—we will distinguish an operator, $\widehat{\mathcal{O}}$ from its matrix representation, $\boldsymbol{\mathcal{O}}$—as well as all of its matrix elements—by introducing the overhead hat symbol and bold-facing the actual symbol for the operator, respectively.

In particular, notice that we can write any arbitrary matrix element—for example, the $\left(j,k\right)$th element—of the adiabatic connection in the basis of the dark states—such as, the $\{\left\vert D_k \right\rangle\}$'s that span the zero-energy-eigenvalue eigenvector subspace—as:
\begin{equation}
\begin{split}                
\sum_{\mu}\left(\bm{A_{\mu}}\right)_{jk}d\kappa_{\mu}&=\sum_{\mu}\left<D_{k}|\widehat{\partial}_{\mu}|D_{j}\right>d\kappa_{\mu}\\
&=\left<D_{k}|\widehat{\nabla}_{\overrightarrow{R}}|D_{j}\right> \cdot d\overrightarrow{R},\\\label{eq:A_me}
\end{split}
\end{equation}
where $\widehat{\partial}_{\mu}=\partial/\partial\kappa_{\mu}$, and  $\widehat{\nabla}_{\overrightarrow{R}}$ and $d\overrightarrow{R}$ are the vector differential operator and the differential line element in the Hamiltonian configuration space, respectively. More specifically, this configuration space is a three-dimensional manifold of the inter-waveguide coupling coefficients that appear in the starting definition of the coupled-mode Hamiltonian: $\widehat{H} = \sum_{\mu}\kappa_{\mu}\widehat{a}_{\mu}\widehat{a}^{\dagger}_{\textrm{C}}+\kappa_{\mu}\widehat{a}^{\dagger}_{\mu}\widehat{a}_{\textrm{C}},$ where $\mu = \{\textrm{E}, \textrm{A}, \textrm{W}\}.$ The coupling strengths between the east, the auxiliary, and the west waveguides and the central waveguide are denoted by $\kappa_{\textrm{E}},$ $\kappa_{\textrm{A}},$ and $\kappa_{\textrm{W}},$ respectively.
Therefore, we can treat this configuration space as analogous to a three-dimensional, locally rectangular Cartesian coordinate space with a locally Euclidean metric.

Consequently, if we make a transformation from rectangular Cartesian coordinates to spherical polar coordinates: $\left(\widetilde r, \widetilde \theta, \widetilde \varphi\right),$ then we can re-write the right-hand side of the above expression as:
\begin{equation}
\begin{split}                
\sum_{\mu}\left(\bm{A_{\mu}}\right)_{jk}d\kappa_{\mu}&=\left<D_{k}|\widehat{\partial}_{\widetilde r}|D_{j}\right>d\widetilde r+\left<D_{k}|\frac{1}{\widetilde r}\widehat{\partial}_{\widetilde \theta}|D_{j}\right>\widetilde r d\widetilde\theta+\left<D_{k}|\frac{1}{\widetilde r\sin\widetilde\theta}\widehat{\partial}_{\widetilde \varphi}|D_{j}\right>\widetilde r\sin\widetilde \theta d\widetilde \varphi\\
&=\left<D_{k}|\widehat{\partial}_{\widetilde r}|D_{j}\right>d\widetilde r+\left<D_{k}|\widehat{\partial}_{\widetilde \theta}|D_{j}\right> d\widetilde\theta+\left<D_{k}|\widehat{\partial}_{\widetilde \varphi}|D_{j}\right> d\widetilde \varphi.\\\label{eq:A_me1}
\end{split}
\end{equation}
To effect one further transformation from the conventional, spherical polar coordinate system to the one that was specifically used in Ref. \cite{neef2023three}, we make the following substitutions: $\widetilde \varphi = \theta, \; \widetilde \theta = \pi/2-\varphi.$ Subsequently, we can write the above expression as:
\begin{equation}
\begin{split}                
\sum_{\mu}\left(\bm{A_{\mu}}\right)_{jk}d\kappa_{\mu}&=\left<D_{k}|\widehat{\partial}_{ r}|D_{j}\right>dr+\left<D_{k}|\widehat{\partial}_{\theta}|D_{j}\right>d\theta+\left<D_{k}|\widehat{\partial}_{ \varphi}|D_{j}\right> d\varphi\\
&=\left(\bm{A_{r}}\right)_{jk}dr+\left(\bm{A_{\theta}}\right)_{jk} d\theta+\left(\bm{A_{\varphi}}\right)_{jk} d\varphi,\\\label{eq:A_me2}
\end{split}
\end{equation}
where $\theta = \tan^{-1}\left(\kappa_{\textrm{A}}/\kappa_{\textrm{E}}\right),$ $\varphi = \tan^{-1}\left(\kappa_{\textrm{W}}/\sqrt{\kappa^2_{\textrm{E}}+\kappa^2_{\textrm{A}}}\right),$ and $r=+\sqrt{\kappa^2_{\textrm{E}}+\kappa^2_{\textrm{A}}+\kappa^2_{\textrm{W}}},$ respectively. Finally, following Ref. \cite{neef2023three}, we define the bosonic creation operators that create the above-mentioned, two-fold, energy-degenerate, dark modes as: 
\begin{equation}
\begin{split}                
\widehat{D}^{\dagger}_1&=\sin\theta\widehat{a}^{\dagger}_{\textrm{E}}-\cos\theta\widehat{a}^{\dagger}_{\textrm{A}}\\
\widehat{D}^{\dagger}_2&=\cos\varphi\widehat{a}^{\dagger}_{\textrm{W}}-\cos\theta\sin\varphi\widehat{a}^{\dagger}_{\textrm{E}}-\sin\theta\sin\varphi\widehat{a}^{\dagger}_{\textrm{A}},\\\label{eq:D1andD2}
\end{split}
\end{equation}
From the above relations, one can readily verify that $\left[\widehat{D}_i,\widehat{D}^\dagger_j\right]=\delta_{ij} \; \forall \; \{i, j\} \in \{1, 2\}$, as the above operators create two independent, uncoupled bosonic modes, and thereby, have canonically conjugate bosonic commutation relations.

\textbf{The key idea behind our derivation is that the matrix representation of the unitary operator, as defined in Eq. (\ref{eq:U_matrix}), can be obtained from the matrix elements defined in Eq. (\ref{eq:A_me2}), and these matrix elements can be computed using commutators and operator algebraic methods, as described below.} To this end, we begin by introducing two additional bosonic operators:
\begin{subequations}
\begin{align}                
\widehat{D}^{\dagger}_3&=\partial_{\theta}\widehat{D}^{\dagger}_1\label{eq:D3}\\
\widehat{D}^{\dagger}_4&=\partial_{\varphi}\widehat{D}^{\dagger}_2,\label{eq:D4}
\end{align}
\end{subequations}
which are the linear transformations of $\widehat{D}^{\dagger}_1$ and $\widehat{D}^{\dagger}_2$, and satisfy the same canonical bosonic commutation relations: $\left[\widehat{D}_i,\widehat{D}^\dagger_j\right]=\delta_{ij} \; \forall \; \{i, j\} \in \{3, 4\}$. From the definitions of these four operators, we can make the following simple, yet useful, observations that we will subsequently use to derive the generalized expressions for the matrix elements of $\widehat{A}_{\theta}$ and $\widehat{A}_{\varphi}$:
\begin{subequations}
\begin{align}                
\partial_{\theta}\widehat{D}^{\dagger}_2&=\sin\varphi\widehat{D}^{\dagger}_1\label{eq:Commut1}\\
\left[\widehat{D}_1,\widehat{D}^{\dagger}_3\right] &= 0\label{eq:Commut2}\\
\left[\widehat{D}_2,\widehat{D}^{\dagger}_3\right] &= -\sin\varphi\label{eq:Commut3}\\
\left[\widehat{D}_2,\widehat{D}^{\dagger}_4\right] &= 0\label{eq:Commut4}
\end{align}
\end{subequations}
For completeness, we also note that this set of operators: $\{\widehat{D}_1, \widehat{D}_2, \widehat{D}_3, \widehat{D}_4\}$ all commute with each other to zero.

We are now well-prepared to commence the actual process of computing the relevant matrix elements. First, we note that, as is evident from the definitions of the operators that create the dark modes, none of the dark states, $\left|D_k\right>$—or, any well-defined, orthonormal basis of states that is constructed from such dark states—have any dependence on the radius, $r.$ Therefore, all matrix elements of the form: $\left(\bm{A_r}\right)_{jk} = \left<D_{k}|\widehat{\partial}_{r}|D_{j}\right>$ are identically zero. Second, the dark states, however, do have non-trivial dependencies on $\theta$ and $\varphi,$ and in the next two subsections, we will derive the exact expressions for $\left<D_{k}|\widehat{\partial}_{\theta}|D_{j}\right>$ and $\left<D_{k}|\widehat{\partial}_{\varphi}|D_{j}\right>,$ respectively.

\subsection{Matrix Elements of the Azimuthal Component of the Gauge Field}\label{sec:Azimuth}
In this subsection, we will derive a generalized expression for matrix elements of the form: $\left<D_{k}|\widehat{\partial}_{\theta}|D_{j}\right>$ that is true for $N \geq 2$. We summarize our conclusions in the statement of the following theorem that we subsequently prove.

\begin{tcolorbox}
\textbf{Theorem 1.}—The matrix representation of the azimuthal component of the non-Abelian gauge field, that is, the $\widehat{A}_{\theta}$ operator—in the basis of the energy-degenerate dark states, and within subspaces in which the total photon number for each of such basis states is conserved—is identical to that of the operator: $-\sin\varphi\left(\widehat{D}^{\dagger}_1\widehat{D}_2-\widehat{D}_1\widehat{D}^{\dagger}_2\right).$ Equivalently, $\bm{A_{\theta}} = -\sin\varphi\left(\bm{D_1}^{\dagger}\bm{D_2}-\bm{D_1}\bm{D_2}^{\dagger}\right)$
\end{tcolorbox}

\emph{Proof.}—We construct and represent our matrices in specific subspaces that are embedded within the zero energy-eigenvalue dark subspace in which the total number of excitations is conserved. For the sake of generality, let us denote this total photon number as $N^\prime$. Later, we will discover that $N^\prime = N-1$, when we will be constructing matrix representations of $N$-dimensional holonomies. Consider the generalized form of the element—of such a matrix—where the $\{\widehat{D}_1, \widehat{D}^{\dagger}_1\}$ and the $\{\widehat{D}_2, \widehat{D}^{\dagger}_2\}$ operators are assumed to act on the first and second entries of the kets and bras, respectively:
\begin{equation}
\begin{split}  
\bm{M_{\theta}} &= \braket{N_{\textrm{B}}, N^\prime-N_{\textrm{B}}|\widehat{\partial}_{\theta}|N_{\textrm{K}}, N^\prime-N_{\textrm{K}}}\\
&= \braket{0,0|\frac{\left(\widehat{D}_2\right)^{N^\prime-N_{\textrm{B}}}\left(\widehat{D}_1\right)^{N_{\textrm{B}}}}{\sqrt{\left(N^\prime-N_{\textrm{B}}\right)! \left(N_{\textrm{B}}\right)! }}|\widehat{\partial}_{\theta}|\frac{\left(\widehat{D}^{\dagger}_1\right)^{N_{\textrm{K}}}\left(\widehat{D}^{\dagger}_2\right)^{N^\prime-N_{\textrm{K}}}}{\sqrt{\left(N_{\textrm{K}}\right)! \left(N^\prime-N_{\textrm{K}}\right)! }}|0,0}\\
&= \frac{1}{\sqrt{\left(N^{\prime}-N_{\textrm{B}}\right)! \left(N_{\textrm{B}}\right)!\left(N_{\textrm{K}}\right)!\left(N^{\prime}-N_{\textrm{K}}\right)!}}\Bigg\{\braket{0,0|\left(\widehat{D}_2\right)^{N^{\prime}-N_{\textrm{B}}}\left(\widehat{D}_1\right)^{N_{\textrm{B}}}\left(\widehat{D}^{\dagger}_1\right)^{N_{\textrm{K}}}\left[\widehat{\partial_{\theta}}\left(\widehat{D}^{\dagger}_2\right)^{N^{\prime}-N_{\textrm{K}}}\right]|0,0}\\ 
&+ 
\braket{0,0|\left(\widehat{D}_2\right)^{N^{\prime}-N_{\textrm{B}}}\left(\widehat{D}_1\right)^{N_{\textrm{B}}}\left[\widehat{\partial}_{\theta}\left(\widehat{D}^{\dagger}_1\right)^{N_{\textrm{K}}}\right]\left(\widehat{D}^{\dagger}_2\right)^{N^{\prime}-N_{\textrm{K}}}|0,0}\Bigg\}\\
&= \bm{M_{\theta}^{\left(1\right)}} + \bm{M_{\theta}^{\left(2\right)}},\\\label{eq:ME_theta}
\end{split}
\end{equation}
where, we have denoted the first and the second terms on the right-hand side as $\bm{M_{\theta}^{\left(1\right)}}$ and $\bm{M_{\theta}^{\left(2\right)}},$ respectively. 
Both $N_{\textrm{K}}$ and $N_{\textrm{B}}$ are assumed to be non-negative integers and range from $0$ to a maximum value of $N^{\prime}$.

Let us first focus on $\bm{M_{\theta}^{\left(1\right)}}$. Using Eq. (\ref{eq:Commut1}), we can further simplify its expression to:
\begin{equation}
\begin{split}
\bm{M_{\theta}^{\left(1\right)}} = \frac{\sin\varphi\left(N^{\prime}-N_{\textrm{K}}\right)}{\sqrt{\left(N^{\prime}-N_{\textrm{B}}\right)! \left(N_{\textrm{B}}\right)!\left(N_{\textrm{K}}\right)! \left(N^{\prime}-N_{\textrm{K}}\right)! }}\braket{0,0|\left(\widehat{D}_2\right)^{N^{\prime}-N_{\textrm{B}}}\left(\widehat{D}_1\right)^{N_{\textrm{B}}}\left(\widehat{D}^{\dagger}_1\right)^{N_{\textrm{K}}+1}\left(\widehat{D}^{\dagger}_2\right)^{N^{\prime}-N_{\textrm{K}}-1}|0,0}.\\\label{eq:ME_theta11} 
\end{split}
\end{equation}
Notice that for $\bm{M_{\theta}^{\left(1\right)}}$ to be non-vanishing, all of the $\widehat{D}_1$'s have to be balanced by an equal number of $\widehat{D}^{\dagger}_1$'s, which implies that $N_{\textrm{B}} = N_{\textrm{K}} + 1$; equivalently, one can draw a similar conclusion by demanding that the $\widehat{D}_2$'s are balanced by an equal number of $\widehat{D}^{\dagger}_2$'s. Therefore, we can further simplify the above expression to:   
\begin{equation}
\begin{split}
\bm{M_{\theta}^{\left(1\right)}} 
&= \frac{\sin\varphi\left(N^{\prime}-N_{\textrm{K}}\right)}{\sqrt{\left(N^{\prime}-N_{\textrm{K}}-1\right)! \left(N_{\textrm{K}}+1\right)!\left(N_{\textrm{K}}\right)! \left(N^{\prime}-N_{\textrm{K}}\right)! }}\braket{0,0|\left(\widehat{D}_2\right)^{N^{\prime}-N_{\textrm{K}}-1}\left(\widehat{D}_1\right)^{N_{\textrm{K}}+1}\left(\widehat{D}^{\dagger}_1\right)^{N_{\textrm{K}}+1}\left(\widehat{D}^{\dagger}_2\right)^{N^{\prime}-N_{\textrm{K}}-1}|0,0}\delta_{N_{\textrm{B}},N_{\textrm{K}}+1}\\
&= \frac{\sin\varphi\sqrt{\left(N^{\prime}-N_{\textrm{K}}\right)\left(N_{\textrm{K}}+1\right)}}{\sqrt{\left[\left(N^{\prime}-N_{\textrm{K}}-1\right)!\right]^2 \left[\left(N_{\textrm{K}}+1\right)!\right]^2 }}\braket{0,0|\left(\widehat{D}_2\right)^{N^{\prime}-N_{\textrm{K}}-1}\left(\widehat{D}_1\right)^{N_{\textrm{K}}+1}\left(\widehat{D}^{\dagger}_1\right)^{N_{\textrm{K}}+1}\left(\widehat{D}^{\dagger}_2\right)^{N^{\prime}-N_{\textrm{K}}-1}|0,0}\delta_{N_{\textrm{B}},N_{\textrm{K}}+1}\\
&= \sin\varphi\sqrt{\left(N^{\prime}-N_{\textrm{K}}\right)\left(N_{\textrm{K}}+1\right)}\braket{N_{\textrm{K}}+1,N^{\prime}-N_{\textrm{K}}-1|N_{\textrm{K}}+1,N^{\prime}-N_{\textrm{K}}-1}\delta_{N_{\textrm{B}},N_{\textrm{K}}+1}\\
&= \sin\varphi\sqrt{\left(N^{\prime}-N_{\textrm{K}}\right)\left(N_{\textrm{K}}+1\right)}\delta_{N_{\textrm{B}},N_{\textrm{K}}+1}\\\label{eq:ME_theta12}\end{split}
\end{equation}

We now turn our attention to simplifying $\bm{M_{\theta}^{\left(2\right)}}$. In particular, notice that $\partial_{\theta}\widehat{D}^{\dagger}_1=\widehat{D}^{\dagger}_3,$ and that $\widehat{D}^{\dagger}_3$ commutes with all of the operators, except for $\widehat{D}_2$, that appear in the expression for $\bm{M_{\theta}^{\left(2\right)}}$—see, for example, Eqs. (\ref{eq:Commut2}) and (\ref{eq:Commut3}). Therefore, we can rearrange the ordering of the operators as follows:
\begin{equation}
\bm{M_{\theta}^{\left(2\right)}} = \frac{N_{\textrm{K}}}{\sqrt{\left(N^{\prime}-N_{\textrm{B}}\right)! \left(N_{\textrm{B}}\right)! \left(N_{\textrm{K}}\right)! \left(N^{\prime}-N_{\textrm{K}}\right)!}}\braket{0,0|\left(\widehat{D}_2\right)^{N^{\prime}-N_{\textrm{B}}}\widehat{D}^{\dagger}_3\left(\widehat{D}_1\right)^{N_{\textrm{B}}}\left(\widehat{D}^{\dagger}_2\right)^{N^{\prime}-N_{\textrm{K}}}\left(\widehat{D}^{\dagger}_1\right)^{N_{\textrm{K}}-1}|0,0}.\label{eq:ME_theta21} 
\end{equation}

We now invoke one particular property of commutators, namely, if $\widehat{A}$ and $\widehat{B}$ are two operators that both commute with their commutator, $\left[\widehat{A}, \widehat{B}\right],$ then, for any positive integer, $n,$ one can show that $\left[\widehat{A}^n, \widehat{B}\right] = n\widehat{A}^{n-1}\left[\widehat{A}, \widehat{B}\right]$; see, for example, Ref. \cite{merzbacher1998quantum}. From this rule and from Eq. (\ref{eq:Commut3}), we arrive at this commutator relation:
\begin{equation}                
\left[\widehat{D}_2^{N^{\prime}-N_{\textrm{B}}},\widehat{D}^{\dagger}_3\right] = \left(N^{\prime}-N_{\textrm{B}}\right)\left(-\sin\varphi\right)\widehat{D}_2^{N^{\prime}-N_{\textrm{B}}-1},\label{eq:Commut5}
\end{equation}
which enables us to break up the expression in Eq. (\ref{eq:ME_theta21}) into two terms:
\begin{equation}
\begin{split}
\bm{M_{\theta}^{\left(2\right)}} &= \frac{N_{\textrm{K}}}{\sqrt{\left(N^{\prime}-N_{\textrm{B}}\right)! \left(N_{\textrm{B}}\right)! \left(N_{\textrm{K}}\right)! \left(N^{\prime}-N_{\textrm{K}}\right)!}}\braket{0,0|\widehat{D}^{\dagger}_3\left(\widehat{D}_2\right)^{N^{\prime}-N_{\textrm{B}}}\left(\widehat{D}_1\right)^{N_{\textrm{B}}}\left(\widehat{D}^{\dagger}_2\right)^{N^{\prime}-N_{\textrm{K}}}\left(\widehat{D}^{\dagger}_1\right)^{N_{\textrm{K}}-1}|0,0}\\
&+ \frac{\left(N_{\textrm{K}}\right)\left(N^{\prime}-N_{\textrm{B}}\right)\left(-\sin\varphi\right)}{\sqrt{\left(N^{\prime}-N_{\textrm{B}}\right)! \left(N_{\textrm{B}}\right)! \left(N_{\textrm{K}}\right)! \left(N^{\prime}-N_{\textrm{K}}\right)!}}\braket{0,0|\left(\widehat{D}_2\right)^{N^{\prime}-N_{\textrm{B}}-1}\left(\widehat{D}_1\right)^{N_{\textrm{B}}}\left(\widehat{D}^{\dagger}_2\right)^{N^{\prime}-N_{\textrm{K}}}\left(\widehat{D}^{\dagger}_1\right)^{N_{\textrm{K}}-1}|0,0}.\\\label{eq:ME_theta21_2}
\end{split}
\end{equation}
Now, the first term in the above expression has to be zero, as $\left\langle 0,0 \right\vert \widehat{D}^{\dagger}_3 = \left(\widehat{D}_3\left\vert 0,0 \right\rangle\right)^{\dagger} = 0$, which stems from the fact that $\widehat{D}^{\dagger}_3$ can only be expressed as a linear combination of bosonic creation operators. Alternatively, if one wishes to remain agnostic with regard to the action of $\widehat{D}^{\dagger}_3$ on the bra $\left\langle 0,0 \right\vert,$ then, consider the action of this product of operators: $\left(\widehat{D}_2\right)^{N^{\prime}-N_{\textrm{B}}}\left(\widehat{D}^{\dagger}_2\right)^{N^{\prime}-N_{\textrm{K}}}\left(\widehat{D}_1\right)^{N_{\textrm{B}}}\left(\widehat{D}^{\dagger}_1\right)^{N_{\textrm{K}}-1}$ on the ket $\left\vert 0,0 \right\rangle.$ There are no possible solutions for $N_{\textrm{K}}$ and $N_{\textrm{B}}$ that will simultaneously ensure that all of the $\widehat{D}_1$'s are balanced by an equal number of $\widehat{D}^{\dagger}_1$'s, and all of the $\widehat{D}_2$'s are balanced by an equal number of $\widehat{D}^{\dagger}_2$'s.

We also note that the second term will be non-zero if and only if the total number of $\widehat{D}_1$'s is equal to the total number of $\widehat{D}^{\dagger}_1$'s, and similarly, if and only if all of the $\widehat{D}_2$'s are balanced by an equal number of $\widehat{D}^{\dagger}_2$'s—each of these requirements are simultaneously satisfied by demanding that: $N_{\textrm{B}} = N_{\textrm{K}}-1.$

As a consequence, Eq. (\ref{eq:ME_theta21_2}) further reduces to:
\begin{equation}
\begin{split}
\bm{M_{\theta}^{\left(2\right)}} &= \frac{\left(N_{\textrm{K}}\right)\left(N^{\prime}-N_{\textrm{K}}+1\right)\left(-\sin\varphi\right)}{\sqrt{\left(N^{\prime}-N_{\textrm{K}}+1\right)! \left(N_{\textrm{K}}-1\right)! \left(N_{\textrm{K}}\right)! \left(N^{\prime}-N_{\textrm{K}}\right)!}}\braket{0,0|\left(\widehat{D}_2\right)^{N^{\prime}-N_{\textrm{K}}}\left(\widehat{D}_1\right)^{N_{\textrm{K}}-1}\left(\widehat{D}^{\dagger}_2\right)^{N^{\prime}-N_{\textrm{K}}}\left(\widehat{D}^{\dagger}_1\right)^{N_{\textrm{K}}-1}|0,0}\delta_{N_{\textrm{B}},N_{\textrm{K}}-1}\\
&= \frac{\sqrt{\left(N_{\textrm{K}}\right)\left(N^{\prime}-N_{\textrm{K}}+1\right)}\left(-\sin\varphi\right)}{\sqrt{\left[\left(N^{\prime}-N_{\textrm{K}}\right)!\right]^2 \left[\left(N_{\textrm{K}}-1\right)!\right]^2}}\braket{0,0|\left(\widehat{D}_2\right)^{N^{\prime}-N_{\textrm{K}}}\left(\widehat{D}_1\right)^{N_{\textrm{K}}-1}\left(\widehat{D}^{\dagger}_2\right)^{N^{\prime}-N_{\textrm{K}}}\left(\widehat{D}^{\dagger}_1\right)^{N_{\textrm{K}}-1}|0,0}\delta_{N_{\textrm{B}},N_{\textrm{K}}-1}\\
&= \sqrt{N_{\textrm{K}}\left(N^{\prime}-N_{\textrm{K}}+1\right)}\left(-\sin\varphi\right)\braket{N_{\textrm{K}}-1,N^{\prime}-N_{\textrm{K}}|N_{\textrm{K}}-1,N^{\prime}-N_{\textrm{K}}}\delta_{N_{\textrm{B}},N_{\textrm{K}}-1}\\
&= \sqrt{N_{\textrm{K}}\left(N^{\prime}-N_{\textrm{K}}+1\right)}\left(-\sin\varphi\right)\delta_{N_{\textrm{B}},N_{\textrm{K}}-1}\\\label{eq:ME_theta22}
\end{split}
\end{equation}
Finally, we can write down the combined expression for $\bm{M_{\theta}}$ as follows:
\begin{equation}
\begin{split}
\bm{M_{\theta}}&= \bm{M_{\theta}^{\left(1\right)}} + \bm{M_{\theta}^{\left(2\right)}}\\
&= \sin\varphi\left(\sqrt{\left(N^{\prime}-N_{\textrm{K}}\right)\left(N_{\textrm{K}}+1\right)}\delta_{N_{\textrm{B}},N_{\textrm{K}}+1} - \sqrt{N_{\textrm{K}}\left(N^{\prime}-N_{\textrm{K}}+1\right)}\delta_{N_{\textrm{B}},N_{\textrm{K}}-1}\right)\\\label{eq:ME_theta_final}
\end{split}
\end{equation}

On the other hand, if we now consider the matrix elements of the operator $-\sin\varphi\left(\widehat{D}^{\dagger}_1\widehat{D}_2-\widehat{D}_1\widehat{D}^{\dagger}_2\right)$ in the basis of the dark states, such as:  
\begin{equation}
\begin{split}  
\bm{\widetilde M_{\theta}} &= \braket{N_{\textrm{K}}, N^{\prime}-N_{\textrm{K}}|-\sin\varphi\left(\widehat{D}^{\dagger}_1\widehat{D}_2-\widehat{D}_1\widehat{D}^{\dagger}_2\right)|N_{\textrm{B}}, N^{\prime}-N_{\textrm{B}}}\\
&= -\sin\varphi\left(\sqrt{N_{\textrm{K}}\left(N^{\prime}-N_{\textrm{K}}+1\right)}\delta_{N_{\textrm{B}},N_{\textrm{K}}-1} - \sqrt{\left(N^{\prime}-N_{\textrm{K}}\right)\left(N_{\textrm{K}}+1\right)}\delta_{N_{\textrm{B}},N_{\textrm{K}}+1}\right)\\
&= \sin\varphi\left(\sqrt{\left(N^{\prime}-N_{\textrm{K}}\right)\left(N_{\textrm{K}}+1\right)}\delta_{N_{\textrm{B}},N_{\textrm{K}}+1} - \sqrt{N_{\textrm{K}}\left(N^{\prime}-N_{\textrm{K}}+1\right)}\delta_{N_{\textrm{B}},N_{\textrm{K}}-1}\right),\\\label{eq:ME_theta_tilde}
\end{split}
\end{equation}
then, from Eqs. (\ref{eq:ME_theta_final}) and (\ref{eq:ME_theta_tilde}), we arrive at the following result:
\begin{equation}
\begin{split}  
\bm{M_{\theta}} &= \braket{N_{\textrm{B}}, N^{\prime}-N_{\textrm{B}}|\widehat{\partial}_{\theta}|N_{\textrm{K}}, N^{\prime}-N_{\textrm{K}}}\\
&= \bm{\widetilde M_{\theta}}\\
&= \braket{N_{\textrm{K}}, N^{\prime}-N_{\textrm{K}}|-\sin\varphi\left(\widehat{D}^{\dagger}_1\widehat{D}_2-\widehat{D}_1\widehat{D}^{\dagger}_2\right)|N_{\textrm{B}}, N^{\prime}-N_{\textrm{B}}}.\\\label{eq:ME_theta_equality1}
\end{split}
\end{equation}
Since the $\left(j,k\right)$th element of the non-Abelian gauge potential is defined as: $\left(\widehat{A}_{\mu}\right)_{jk}=\braket{D_{k}|\widehat{\partial}_{\mu}|D_{j}},$ that is, the ket and the bra are interchanged as compared with the usual definition of the matrix element—see, for example, Eqs. (7) and (11) of Ref. \cite{PhysRevLett.52.2111}—we can deduce from the above equation that:
\begin{equation}
\begin{split}  
\braket{N_{\textrm{K}}, N^{\prime}-N_{\textrm{K}}|\widehat{A}_{\theta}|N_{\textrm{B}}, N^{\prime}-N_{\textrm{B}}} &= \braket{N_{\textrm{B}}, N^{\prime}-N_{\textrm{B}}|\widehat{\partial}_{\theta}|N_{\textrm{K}}, N^{\prime}-N_{\textrm{K}}}\\
&= \braket{N_{\textrm{K}}, N^{\prime}-N_{\textrm{K}}|-\sin\varphi\left(\widehat{D}^{\dagger}_1\widehat{D}_2-\widehat{D}_1\widehat{D}^{\dagger}_2\right)|N_{\textrm{B}}, N^{\prime}-N_{\textrm{B}}}.\\\label{eq:ME_theta_equality2}
\end{split}
\end{equation}

Therefore, in the basis of dark states and within subspaces in which the total photon number is conserved, the matrix representations of the operators $\widehat{A}_{\theta}$ and $-\sin\varphi\left(\widehat{D}^{\dagger}_1\widehat{D}_2-\widehat{D}_1\widehat{D}^{\dagger}_2\right)$ are identical. \hfill $\qedsymbol{}$

\subsection{Matrix Elements of the Polar Component of the Gauge Field}\label{sec:Polar}

In this subsection, we will show that the matrix representation of the operator, $\widehat{A}_{\varphi}$ is the zero matrix, regardless of the dimension of the holonomy. We summarize our conclusions in the statement of the following theorem, and subsequently, furnish its proof.

\begin{tcolorbox}
\textbf{Theorem 2.}—All of the elements of the matrix representation of the polar component of the non-Abelian gauge field, that is, the $\widehat{A_{\varphi}}$ operator—in the basis of the energy-degenerate dark states and within subspaces in which the total photon number for each of such basis states is conserved—are zero. Equivalently, $\bm{A_{\varphi}} = \bm{0}$. 
\end{tcolorbox}

\emph{Proof.}—Let us consider—just like we did for the proof in the preceding subsection—a generalized matrix element of the following form, where the $\{\widehat{D}_1, \widehat{D}^{\dagger}_1\}$ and the $\{\widehat{D}_2, \widehat{D}^{\dagger}_2\}$ operators are assumed to act on the first and second entries of the kets and bras, respectively:
\begin{equation}
\begin{split}  
\bm{M_{\varphi}} &= \braket{N_{\textrm{B}}, N^{\prime}-N_{\textrm{B}}|\widehat{\partial}_{\varphi}|N_{\textrm{K}}, N^{\prime}-N_{\textrm{K}}}\\
&= \braket{0,0|\frac{\left(\widehat{D}_2\right)^{N^{\prime}-N_{\textrm{B}}}\left(\widehat{D}_1\right)^{N_{\textrm{B}}}}{\sqrt{\left(N^{\prime}-N_{\textrm{B}}\right)! \left(N_{\textrm{B}}\right)! }}|\widehat{\partial}_{\varphi}|\frac{\left(\widehat{D}^{\dagger}_1\right)^{N_{\textrm{K}}}\left(\widehat{D}^{\dagger}_2\right)^{N^{\prime}-N_{\textrm{K}}}}{\sqrt{\left(N_{\textrm{K}}\right)! \left(N^{\prime}-N_{\textrm{K}}\right)! }}|0,0}\\
&= \braket{0,0|\frac{\left(\widehat{D}_2\right)^{N^{\prime}-N_{\textrm{B}}}\left(\widehat{D}_1\right)^{N_{\textrm{B}}}}{\sqrt{\left(N^{\prime}-N_{\textrm{B}}\right)! \left(N_{\textrm{B}}\right)! }} \times \frac{\left(\widehat{D}^{\dagger}_1\right)^{N_{\textrm{K}}}\widehat{\partial}_{\varphi}\left(\widehat{D}^{\dagger}_2\right)^{N^{\prime}-N_{\textrm{K}}}}{\sqrt{\left(N_{\textrm{K}}\right)! \left(N^{\prime}-N_{\textrm{K}}\right)! }}|0,0}. \\\label{eq:ME_varphi}
\end{split}
\end{equation}
To arrive at the last line of the above expression, we have utilized the fact that $\partial_{\varphi}\widehat{D}^{\dagger}_1 = 0.$

Since the sets of operators $\{\widehat{D}_1,\widehat{D}^{\dagger}_1\}$ and $\{\widehat{D}_2,\widehat{D}^{\dagger}_2\}$ operate on mutually orthogonal subspaces, therefore, with respect to these two sets of operators, the vacuum can be decomposed as $\left\vert 0,0 \right\rangle = \left\vert 0 \right\rangle_{1} \otimes \left\vert 0 \right\rangle_{2}$. Consequently, we can write the above matrix element as a product of two reduced matrix elements: $\bm{M_{\varphi}} = \bm{M_{\varphi}^{\left(1\right)}M_{\varphi}^{\left(2\right)}},$ where 
\begin{equation}
\begin{split}  
\bm{M_{\varphi}^{\left(1\right)}} &= \braket{N_{\textrm{B}}|\widehat{\partial}_{\varphi}|N_{\textrm{K}}}\\
&= \braket{0|\frac{\left(\widehat{D}_1\right)^{N_{\textrm{B}}}}{\sqrt{\left(N_{\textrm{B}}\right)! }}\frac{\left(\widehat{D}^{\dagger}_1\right)^{N_{\textrm{K}}}}{\sqrt{\left(N_{\textrm{K}}\right)! }}|0},\\\label{eq:ME_varphi1}
\end{split}
\end{equation}
\begin{equation}
\begin{split}  
\bm{M_{\varphi}^{\left(2\right)}} &= \braket{N^{\prime}-N_{\textrm{B}}|\widehat{\partial}_{\varphi}|N^{\prime}-N_{\textrm{K}}}\\
&= \braket{0|\frac{\left(\widehat{D}_2\right)^{N^{\prime}-N_{\textrm{B}}}}{\sqrt{\left(N^{\prime}-N_{\textrm{B}}\right)! }}\frac{\partial_{\varphi}\left(\widehat{D}^{\dagger}_2\right)^{N^{\prime}-N_{\textrm{K}}}}{\sqrt{\left(N^{\prime}-N_{\textrm{K}}\right)! }}|0}\\
&= \left(N^{\prime}-N_{\textrm{K}}\right)\braket{0|\frac{\left(\widehat{D}_2\right)^{N^{\prime}-N_{\textrm{B}}}}{\sqrt{\left(N^{\prime}-N_{\textrm{B}}\right)! }}\frac{\left(\widehat{D}^{\dagger}_2\right)^{N^{\prime}-N_{\textrm{K}}-1}\widehat{D}^{\dagger}_4}{\sqrt{\left(N^{\prime}-N_{\textrm{K}}\right)! }}|0},\\\label{eq:ME_varphi2}
\end{split}
\end{equation}
and we have used the definition of $\widehat{D}^{\dagger}_4$ from Eq. (\ref{eq:D4}) in the above expression for $\bm{M_{\varphi}^{\left(2\right)}}.$
Notice that $\bm{M_{\varphi}^{\left(1\right)}} \neq 0$ if and only if $N_{\textrm{B}} = N_{\textrm{K}}$; however, this equality implies that: 
\begin{equation}
\begin{split}  
\bm{M_{\varphi}^{\left(2\right)}} 
&= \left(N^{\prime}-N_{\textrm{K}}\right)\braket{0|\frac{\left(\widehat{D}_2\right)^{N^{\prime}-N_{\textrm{B}}}}{\sqrt{\left(N^{\prime}-N_{\textrm{B}}\right)! }}\frac{\left(\widehat{D}^{\dagger}_2\right)^{N^{\prime}-N_{\textrm{B}}-1}\widehat{D}^{\dagger}_4}{\sqrt{\left(N^{\prime}-N_{\textrm{B}}\right)! }}|0}.\\\label{eq:ME_varphi21}
\end{split}
\end{equation}

Furthermore, since $\left[\widehat{D}^{\dagger}_2, \widehat{D}^{\dagger}_4\right] = \left[\widehat{D}_2, \widehat{D}^{\dagger}_4\right] = 0$—see, for example, Eq. (\ref{eq:Commut4})—we can recast the above expression for $\bm{M_{\varphi}^{\left(2\right)}}$ as follows:
\begin{equation}
\begin{split}  
\bm{M_{\varphi}^{\left(2\right)}} 
&= \left(N^{\prime}-N_{\textrm{K}}\right)\braket{0|\frac{\widehat{D}^{\dagger}_4\left(\widehat{D}_2\right)^{N^{\prime}-N_{\textrm{B}}}}{\sqrt{\left(N^{\prime}-N_{\textrm{B}}\right)! }}\frac{\left(\widehat{D}^{\dagger}_2\right)^{N^{\prime}-N_{\textrm{B}}-1}}{\sqrt{\left(N^{\prime}-N_{\textrm{B}}\right)! }}|0}.\\\label{eq:ME_varphi22}
\end{split}
\end{equation}
Now, $\left\langle 0 \right\vert \widehat{D}^{\dagger}_4 = \left(\widehat{D}_4\left\vert 0 \right\rangle\right)^{\dagger} = 0,$ as $\widehat{D}^{\dagger}_4$ can only be expressed as a linear combination of bosonic creation operators. Alternatively, the action of this particular product of operators: $\left(\widehat{D}_2\right)^{N^{\prime}-N_{\textrm{B}}}\left(\widehat{D}^{\dagger}_2\right)^{N^{\prime}-N_{\textrm{B}}-1}$ on the vacuum state has to be zero, since there is one additional annihilation operator. Notice that, previously, we had furnished an analogous argument as to why the first term of Eq. (\ref{eq:ME_theta21_2}) should vanish.

Therefore, the product $\bm{M_{\varphi}} = \bm{M_{\varphi}^{\left(1\right)} M_{\varphi}^{\left(2\right)}} = 0,$ and hence, all matrix elements of the $\widehat{\partial}_{\varphi}$ operator—or, equivalently, of the $\widehat{A}_{\varphi}$ operator—are zero in the representation of dark states. \hfill $\qedsymbol{}$

\subsection{The Appearance of the Non-Abelian, Holonomic Phase }\label{sec:Holonomic_Phase}

According to Theorems 1 and 2, the expression for the \textbf{matrix} representation of the unitary holonomy—as defined by Eqs. (\ref{eq:U_matrix}) and (\ref{eq:A_me2})—can be highly simplified in the following manner: 
\begin{equation}
\begin{split}
\bm{U\left(\gamma\right)} &= \mathcal{P}\textrm{exp}\left(\oint_{\gamma}\sum_{\mu}\bm{A_{\mu}}d\kappa_{\mu}\right)\\
&= \mathcal{P}\textrm{exp}\left(\oint_{\gamma}\bm{A_{\theta}}d\theta\right)\\
&= \mathcal{P}\textrm{exp}\left[-\oint_{\gamma}\sin\varphi\left(\bm{D_1}^{\dagger}\bm{D_2}-\bm{D_1}\bm{D_2}^{\dagger}\right)d\theta\right]\\
&= \textrm{exp}\left[-\left(\bm{D_1}^{\dagger}\bm{D_2}-\bm{D_1}\bm{D_2}^{\dagger}\right)\mathcal{P}\oint_{\gamma}\sin\varphi d\theta\right]\\
&= \textrm{exp}\left[-\phi\left(\gamma\right)\left(\bm{D_1}^{\dagger}\bm{D_2}-\bm{D_1}\bm{D_2}^{\dagger}\right)\right],\\
\label{eq:U_matrix1}
\end{split}
\end{equation}
where
\begin{equation}
\phi\left(\gamma\right) = \mathcal{P}\oint_{\gamma}\sin\varphi d\theta \label{eq:NAH_phase}
\end{equation}
is the holonomic phase that is accrued as the input state vector adiabatically propagates along the closed loop, $\gamma$, for one complete closed cycle of traversal.

\textbf{The above algebraic manipulation, which involves identifying and bringing the phase out of the integral, is one of the key steps in our derivation, and is legitimized by the fact that both the left-hand and the right-hand sides of Eq. (\ref{eq:U_matrix1}) are matrices in the bases of dark states.} Specifically, $\bm{D_1}^{\dagger}\bm{D_2}-\bm{D_1}\bm{D_2}^{\dagger}$ is the matrix representation of the corresponding linear operator—analogous to the non-diagonal, coupling terms in the Hamiltonian for two coupled oscillators when represented in the basis of two-mode Fock states—\textbf{and this features enables us to bring this skew-symmetric matrix out of the integral.}

\begin{figure}
\includegraphics[width=\linewidth]{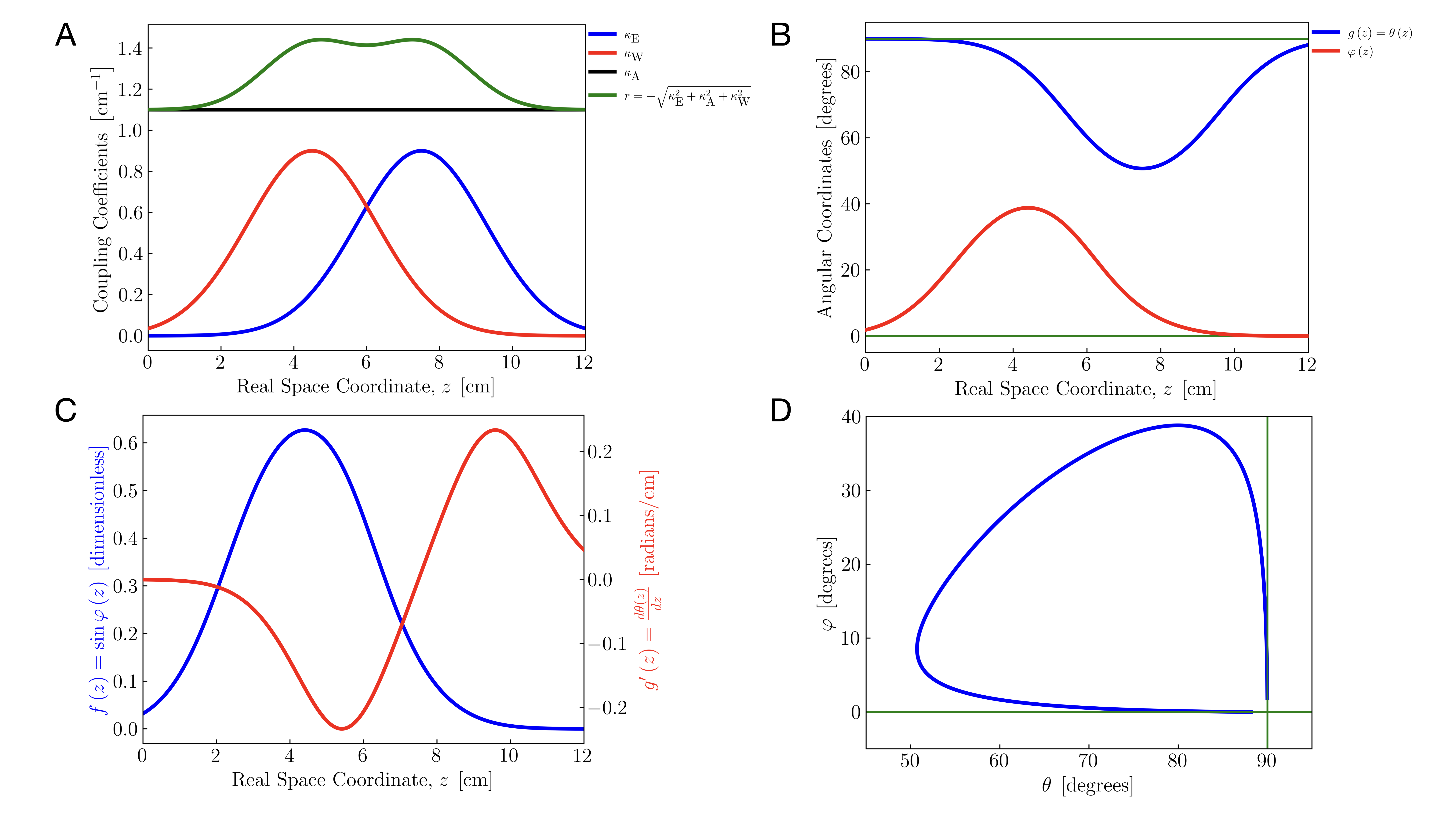}
\caption{\label{fig:fS1} \textbf{Evaluating the phase integral as a closed line integral.} \textbf{A.} The spatial variations in the three inter-waveguide coupling coefficients—and the variation in $r = + \sqrt{\kappa_{\textrm{E}}^2 + \kappa_{\textrm{A}}^2 + \kappa_{\textrm{W}}^2}$, which determines the energy eigenvalues of the bright modes—as functions of the real space coordinate, $z$, due to adiabatic propagation through the holonomic device. \textbf{B.} The spatial variations in the angular coordinates, $\theta$ and $\varphi$—that constitute the coordinate system that was adopted in Ref. \cite{neef2023three}—as functions of the real space coordinate, $z$, due to adiabatic propagation through the holonomic device. The solid, horizontal, green-colored lines indicate the angles of $0^{\circ}$ and $90^{\circ}$—which effectively demarcate the range of values of the angular coordinates. \textbf{C.} The resultant spatial variations in the two integrands that appear in Eqs. (\ref{eq:phase_integral1}) and (\ref{eq:phase_integral5}). \textbf{D.} The trajectory of the input state vector in the $\left(\theta, \varphi\right)$ phase space, as it propagates through the holonomic device, along the real space coordinate, $z$. The intersection of the vertical and the horizontal green-colored, straight lines is supposed to approximate the starting and the ending points of this trajectory. When viewed from the perspective of the reader looking down on the plane of the page, the contour appears have to counterclockwise orientation; conversely, when viewed from the origin of the $\left(r, \theta, \varphi\right)$ coordinate system, this same closed curve has clockwise orientation.}
\end{figure}

The closed line integral above can be evaluated as an integral with respect to the real-space spatial coordinate, $z$ along the length of the quantum photonic, holonomic device:
\begin{equation}
\phi\left(\gamma\right) = \mathcal{P}\oint_{\gamma}\sin\varphi d\theta = \int_{z_i}^{z_f}f(z)g^{\prime}(z) dz,\label{eq:phase_integral1}
\end{equation}
where $f(z) = \sin \varphi(z)$ and $g(z) = \theta(z)$, and $z_i$ and $z_f$ indicate the input and the output facets of the device, respectively. To elucidate the appearance of the holonomic phase from the evaluation of the above phase integral, we adopt the following specific instantiation of the parameterization $\gamma$ that was used to experimentally realize the holonomic device in Ref. \cite{neef2023three}:       
\begin{equation}
\begin{split}
\kappa_{\textrm{E}}\left(z\right) &= \Omega\textrm{exp}\left[-\frac{\left(z-\Bar{z}-\tau\right)^2}{T^2}\right]\\
\kappa_{\textrm{W}}\left(z\right) &= \Omega\textrm{exp}\left[-\frac{\left(z-\Bar{z}+\tau\right)^2}{T^2}\right]\\
\kappa_{\textrm{A}}\left(z\right) &= \kappa_{\textrm{A}}\\
z &\in \left[z_i, z_f\right] = \left[0, 2\Bar{z}\right],\\
\label{eq:phase_integral2}
\end{split}
\end{equation}
which, in turn, implies that:
\begin{equation}
\begin{split}
f(z) &=\sin \varphi(z)\\
&=\sin \left[\textrm{tan}^{-1}\left(\kappa_{\textrm{W}}\Big/{\sqrt{\kappa_{\textrm{E}}^2 + \kappa_{\textrm{A}}^2}}\right)\right]\\
&= \sin\bm{\Bigg[}\textrm{tan}^{-1}\bm{\Bigg(}\Omega\textrm{exp}\left[-\frac{\left(z-\Bar{z}+\tau\right)^2}{T^2}\right]\Bigg/\Bigg\{\Omega^2\textrm{exp}\left[-2\frac{\left(z-\Bar{z}-\tau\right)^2}{T^2}\right]+\kappa_{\textrm{A}}^2\Bigg\}^{1/2}\bm{\Bigg)}\bm{\Bigg]},\\
\label{eq:phase_integral3}
\end{split}
\end{equation}
and that:
\begin{equation}
\begin{split}
g(z) &= \theta(z)\\
&= \textrm{tan}^{-1}\left(\kappa_{\textrm{A}}\Big/\kappa_{\textrm{E}}\left(z\right)\right)\\
&=\textrm{tan}^{-1}\Bigg\{\frac{\kappa_{\textrm{A}}}{\Omega}\textrm{exp}\left[\frac{\left(z-\Bar{z}-\tau\right)^2}{T^2}\right]\Bigg\}.\\
\label{eq:phase_integral4}
\end{split}
\end{equation}

Specifically, the values of the parameters that were used in the experiment, as reported in Ref. \cite{neef2023three}, are: $\kappa_{\textrm{A}} = 1.1\;\textrm{cm}^{-1}, \; \Omega = 0.9\;\textrm{cm}^{-1}, \; \tau = 1.5\;\textrm{cm}, \; T = 2.5\; \textrm{cm}, \; \textrm{and} \; \Bar{z} = 6 \; \textrm{cm}$. The spatial variations in the parameters that define the closed path, $\gamma$—and, as a consequence, those of the integrands in Eq. (\ref{eq:phase_integral1})—as functions of the real-space coordinate—for the above experimentally realized values—are shown in Fig. \ref{fig:fS1}. For this situation, we discover that: \begin{equation}
\phi\left(\gamma\right) = \mathcal{P}\oint_{\gamma}\sin\varphi d\theta = -\int_{0}^{12}f(z)g^{\prime}(z) dz \approx -(-0.33) \; \textrm{rad},\label{eq:phase_integral5}
\end{equation} 
where we have explicitly shown the negative sign that originates from the fact that the closed contour—see, for example, the panel labelled D in Fig. \ref{fig:fS1}—has clockwise orientation when viewed from the perspective of the origin of the $\left(r, \theta, \varphi\right)$ coordinate system. $\phi\left(\gamma\right) \approx 0.33  \; \textrm{rad}$ is in excellent agreement with the value reported in Ref. \cite{neef2023three}.

\subsection{Equivalence of Matrix Representations of Quantum Holonomies and those of Rotation Operators }\label{sec:angular_momentum}
In the final subsection of this section, we establish a formal analogy between the algebraic theory of angular momenta and the underlying nature of real, orthonormal transformations that result from photonic quantum holonomies. We present our central result as a theorem followed by a few, useful corollaries that are readily deducible from this theorem.

\begin{tcolorbox}
\textbf{Theorem 3.}—\textbf{The Representation of non-Abelian Holonomies Theorem}—The $N \times N$ real, orthonormal matrix representation of the quantum holonomy—where the dimension of the holonomy, $N$ is an integer strictly greater than 1—in the basis of the energy-degenerate dark states is equivalent to the $j = \left(N-1\right)/2$—or, the $\left(2j+1\right)$-dimensional—irreducible representation of the rotation operator, $\widehat{\mathscr{D}}\left(\alpha, \beta, \gamma\right) = \widehat{\mathscr{D}}_z\left(\alpha\right)\widehat{\mathscr{D}}_y\left(\beta\right)\widehat{\mathscr{D}}_z\left( \gamma\right)$, where $\alpha$, $\beta$, and $\gamma$ are the three Euler angles such that $\alpha = \gamma = 0$, and $j$ is the angular momentum quantum number.   
\end{tcolorbox}

\emph{Proof.}—In accordance with the usual quantum-mechanical treatment of rotations and of angular momenta—see, for example, Ref. \cite{sakurai2020modern}—the rotation operator acting on the state vectors in the appropriate ket space can be written as:
\begin{equation}
\widehat{\mathscr{D}}\left(\alpha, \beta, \gamma\right) = \widehat{\mathscr{D}}_z\left(\alpha\right)\widehat{\mathscr{D}}_y\left(\beta\right)\widehat{\mathscr{D}}_z\left( \gamma\right),\label{eq:rotation1}
\end{equation}
where $\alpha,$ $\beta,$ and $\gamma$ are the corresponding Euler angles. In particular, in this formalism, the only non-trivial rotation is the one about the $y$-axis, as rotations about the $z$-axis impart an overall, pure phase to the matrix elements of the rotation operator in the basis of $\left|j,m\right\rangle$, which are the eigenstates of the operator corresponding to the $z$-component of the angular momentum, $\widehat{J_z}$. Consequently, we can re-write the above equation as:  \begin{equation}
\widehat{\mathscr{D}}\left(\alpha, \beta, \gamma\right)\bigg\vert_{\alpha = \gamma =0} = \widehat{\mathscr{D}}_y\left(\beta\right) = \textrm{exp}\left(-\frac{i\widehat{J}_y\beta}{\hbar}\right),\label{eq:rotation2}
\end{equation}
where $\widehat{J}_y$ is the operator corresponding to the y-component of the angular momentum. In accordance with Schwinger's oscillator model of angular momentum—see, for example, \cite{schwinger1952, sakurai2020modern}—which establishes a formal analogy between the operator algebra of angular momentum and that of a pair of uncoupled oscillators, we introduce the bosonic operators: $\{\widehat{a}_{+}, \widehat{a}^{\dagger}_{+}\}$ and $\{\widehat{a}_{-}, \widehat{a}^{\dagger}_{-}\}$ corresponding to these two uncoupled, simple harmonic oscillators that we have called, in accordance with the textbook of Sakurai and Napolitano \cite{sakurai2020modern}, the plus-type and the minus-type, respectively. If we substitute the definition of $\widehat{J}_y$ in Schwinger's formalism:  
\begin{equation}
\widehat{J}_y = \frac{\hbar}{2i}\left(\widehat{a}^{\dagger}_{+}\widehat{a}_{-}-\widehat{a}^{\dagger}_{-}\widehat{a}_{+}\right),\label{eq:schwinger1}
\end{equation}
in Eq. (\ref{eq:rotation2}), then we obtain the expression: 
\begin{equation}
\widehat{\mathscr{D}}_y\left(\beta\right) = \textrm{exp}\left[-\frac{\beta}{2}\left(\widehat{a}^{\dagger}_{+}\widehat{a}_{-}-\widehat{a}^{\dagger}_{-}\widehat{a}_{+}\right)\right].\label{eq:schwinger2}
\end{equation}

Let us now represent the above operator as a matrix in the basis of two-mode Fock states, $\left|N_{+}, N_{-}\right\rangle$—where $N_{+}$ and $N_{-}$ are the eigenvalues of the number operators, $\widehat{N}_{+} = \widehat{a}^{\dagger}_{+}\widehat{a}_{+}$ and $\widehat{N}_{-} = \widehat{a}^{\dagger}_{-}\widehat{a}_{-}$, respectively—such that $N_{+} + N_{-} = \left(N-1\right)$, that is, a constant. More specifically, according to Schwinger's formalism, $N=2j+1$, where $j$ is the angular momentum quantum number. We further make the following substitutions—analogous to a change of variables:
\begin{equation}
\begin{split}
\phi\left(\gamma\right) &= \beta/2\\
\widehat{D}_1 &= \widehat{a}_{+}\\
\widehat{D}_2 &= \widehat{a}_{-},\\
\label{eq:schwinger3}
\end{split}
\end{equation}
where $\phi\left(\gamma\right)$, and $\widehat{D}_1$ and $\widehat{D}_2$ are defined by Eqs. (\ref{eq:NAH_phase}) and (\ref{eq:D1andD2}), respectively. The suitability of these substitutions is justified by the following reasons: First, both $\phi\left(\gamma\right)$ and $\beta$ are proper, linear angles; and second, the sets of operators: $\{\widehat{D}_1, \widehat{D}^{\dagger}_1\}$ and $\{\widehat{D}_2, \widehat{D}^{\dagger}_2\}$ similarly correspond to two mutually uncoupled harmonic oscillators, that is, $\left[\widehat{D}_1, \widehat{D}^{\dagger}_2\right] = \left[\widehat{D}_2, \widehat{D}^{\dagger}_1\right] = 0.$ As a consequence, we can write the operator equation, Eq. (\ref{eq:schwinger2}) in the matrix representation as:
\begin{equation}
\pmb{\mathscr{D}_y\left(\phi\left(\gamma\right)\right)} = \textrm{exp}\left[-\phi\left(\gamma\right)\left(\bm{D_1}^{\dagger}\bm{D_2}-\bm{D_1}\bm{D_2}^{\dagger}\right)\right].\label{eq:schwinger4}
\end{equation}
Notice that the matrix on the right-hand side of the above expression is equivalent to the matrix representation of the unitary holonomy matrix in the basis of dark states—see, for example, Eq. (\ref{eq:U_matrix1}); moreover, the set of basis vectors corresponding to the dark states is equal to the set of the basis vectors corresponding to the two-mode Fock states above, that is, $\left|N_{+}, N_{-}\right\rangle$. Therefore,
\begin{equation}
\pmb{\mathscr{D}_y\left(\phi\left(\gamma\right)\right)} = \bm{U\left(\gamma\right)}.\label{eq:theorem3}
\end{equation}
We emphasize that the above equality is merely a statement of equivalence of the matrix representations of the rotation operator and the holonomy in the basis of two-mode Fock states—in which the total photon number is conserved—and not necessarily that of the two operators. The requirement for operating in subspaces in which the total photon number is constant is analogous to—and consistent with—that for constructing rotation matrices that are characterized by a definite value of $j$. \hfill $\qedsymbol{}$

We conclude this subsection by stating a few, useful corollaries that immediately follow as consequences of Theorems 1, 2, and 3:

\textbf{Corollary 1.1}—The matrix representation of the $y$-component of the angular momentum operator and that of the azimuthal component of the non-Abelian gauge field—in the basis of two-mode Fock states in which the total photon number is conserved—are equivalent modulo an overall multiplicative factor.

\emph{Proof.}—From the statement of Theorem 1, and from Eqs. (\ref{eq:schwinger4}) and (\ref{eq:theorem3}), it immediately follows that:  
\begin{equation}
\bm{J_y} = \frac{i\hbar}{2\sin\varphi}\bm{A_{\theta}}.\label{eq:coroll1}
\end{equation} \hfill \qedsymbol{}

\textbf{Corollary 1.2}—Wigner's explicit formula for rotation matrices can be utilized to compute all matrix elements of the holonomy—in the basis of two-mode Fock states in which the total photon number is conserved—for any arbitrary dimension of the holonomy, $N \geq 2$.

\emph{Proof.}—Let us denote, as is usually done, the matrix elements of the rotation operator in the basis of eigen-kets of $\widehat{J}_z$, that is, $\left|j,m\right\rangle$, for an arbitrary, definite value of $j$:  
\begin{equation}
\pmb{\mathscr{D}^{\left(j\right)}_{m^\prime m}\left(\alpha, \beta, \gamma\right)}\bigg\vert_{\alpha=\gamma=0}=\pmb{d^{\left(j\right)}_{m^\prime m}\left(\beta\right)} \equiv \left\langle j, m^{\prime} \right \vert \textrm{exp} \left(\frac{-i\widehat{J}_y \beta}{\hbar}\right)\left\vert j,m \right\rangle.\label{eq:coroll2}
\end{equation}
According to Theorem 3, the above matrix elements are identical to the elements of the matrix representation of the quantum holonomy in the basis of the dark states, assuming that we perform: 1) the change of variables: $\beta = 2\phi\left(\gamma\right)$; and 2) effect the following mapping between the basis vectors: $\left\vert j,m\right\rangle \longmapsto \left\vert N_{+}, N_{-}\right\rangle$, where $j=\left(N_{+}+N_{-}\right)/2$ and $m=\left(N_{+}-N_{-}\right)/2$. This mathematical procedure can be used to construct matrices for a fixed value of $j$, and consequently, for a given, constant dimension of the holonomy, $N = 2j+1$.

Convention dictates that the column vectors, $\left\vert j,m \right\rangle$ of these rotation matrices are sequentially arranged from left to right in such a way that the $m$'s are in a descending order from $+j$ to $-j$. To bring these matrices into exact correspondence with the holonomy matrices—see, for example, Ref. \cite{neef2023three} for two- and three-dimensional instances of such holonomy matrices—the column vectors, $\left\vert N_{+}, N_{-} \right\rangle$ have to be sequentially arranged so that the $N_{+}$'s are in a descending order from $\left(N-1\right)$ to $0$, and conversely, the $N_{-}$'s are in an ascending order from $0$ to $\left(N-1\right)$, while traversing from left to right. The sum $N_{+}+N_{-} = 2j = N-1$ is the total number of photons, which is a conserved quantity for a given dimension of the holonomy.\hfill \qedsymbol{}

\textbf{Corollary 1.3}—A serial cascading—such that the output waveguide facets of the preceding device are directly connected to the corresponding input facets of the following device—of quantum photonic, holonomic devices can be utilized to generate a cumulative, larger value—in an additive manner—of the overall, non-Abelian holonomic phase. We assume that each of these devices are being operated in the same holonomic dimension.

\emph{Proof.}—Equation (\ref{eq:U_matrix1})—which is a direct consequence of Theorems 1 and 2, besides being essential for the correctness of Theorem 3—gives the overall, unitary transformation that is underwent by a quantum state vector—that is presented at the input facets of a quantum photonic, holonomic device—as it adiabatically propagates through the entire length of the device. If $N$ such devices are cascaded in series, so that $\bm{U^{\left(i\right)}\left(\gamma\right)}$ and $\phi^{\left(i\right)}\left(\gamma\right)$ are the unitary transformation matrix of—and the accrued, non-Abelian phase due to propagation through—the $i$-th device, respectively, then the unitary transformation matrix describing the entire series of devices is:
\begin{equation}
\begin{split}
\bm{U^{\left(\textrm{R}\right)}\left(\gamma\right)}&=\prod_{i=1}^{N}\bm{U^{\left(i\right)}\left(\gamma\right)}\\
&=\prod_{i=1}^{N}\textrm{exp}\left[-\phi^{\left(i\right)}\left(\gamma\right)\left(\bm{D_1}^{\dagger}\bm{D_2}-\bm{D_1}\bm{D_2}^{\dagger}\right)\right]\\
&=\textrm{exp}\left[-\left(\sum_{i=1}^{N}\phi^{\left(i\right)}\left(\gamma\right)\right)\left(\bm{D_1}^{\dagger}\bm{D_2}-\bm{D_1}\bm{D_2}^{\dagger}\right)\right]\\
&=\textrm{exp}\left[-\phi^{\left(\textrm{R}\right)}\left(\gamma\right)\left(\bm{D_1}^{\dagger}\bm{D_2}-\bm{D_1}\bm{D_2}^{\dagger}\right)\right].\\
\label{eq:coroll3}
\end{split}
\end{equation}
Notice that in the above algebraic rearrangements, we have utilized the fact that $\bm{D_1}^{\dagger}\bm{D_2}-\bm{D_1}\bm{D_2}^{\dagger}$ is a matrix in the basis of the dark states and this matrix is identical for each of these devices, even though the corresponding operators individually inhabit distinct Hilbert spaces. We identify $\phi^{\left(\textrm{R}\right)}$ as the overall, non-Abelian phase that—by virtue of the above-described additive effect—can be greatly increased. \hfill \qedsymbol{}

\section{Verifying and Characterizing Pure State Entanglement using the R\'enyi Entropy}\label{sec:Renyi}

In the main text, we explored the formation of entanglement as a function of the accumulated, overall, non-Abelian phase; see, for example, Fig. 2 of the main text. For completeness, let us also examine the growth of pure-state entanglement through the lens of the R\'enyi entanglement entropy \cite{RevModPhys.81.865, PhysRevA.54.1838}.

All separable quantum states obey the following entropic inequalities:
\begin{equation}                
\textrm{Tr}\left(\rho_{\textrm{E}}^2\right) \geq \textrm{Tr}\left({\rho}^2\right); \; \textrm{Tr}\left(\rho_{\textrm{W}}^2\right) \geq \textrm{Tr}\left({\rho}^2\right)\label{eq:inequal},
\end{equation}
where $\textrm{Tr}\left(\rho_{\textrm{E}}\right)$ and $\textrm{Tr}\left(\rho_{\textrm{W}}\right)$ are the reduced density matrices of the two pertinent subsystems—which are, in our case, the east and west waveguides, respectively—and $\rho$ is the overall, system density matrix \cite{RevModPhys.81.865, PhysRevA.54.1838, PhysRevLett.95.240407}. More importantly, this method of verifying entanglement—involving explicitly nonlinear functions of the partial density operators—have been proven to be more universally applicable and reliable than all of the Bell-Clauser-Horne-Shimony-Holt (CHSH) inequalities \cite{horodecki1996quantum, PhysRevLett.23.880}. Furthermore, the violation of these entropic inequalities can be also construed as being equivalent to the demonstration of a nonlinear entanglement witness.

\begin{figure}
\includegraphics[width=\linewidth]{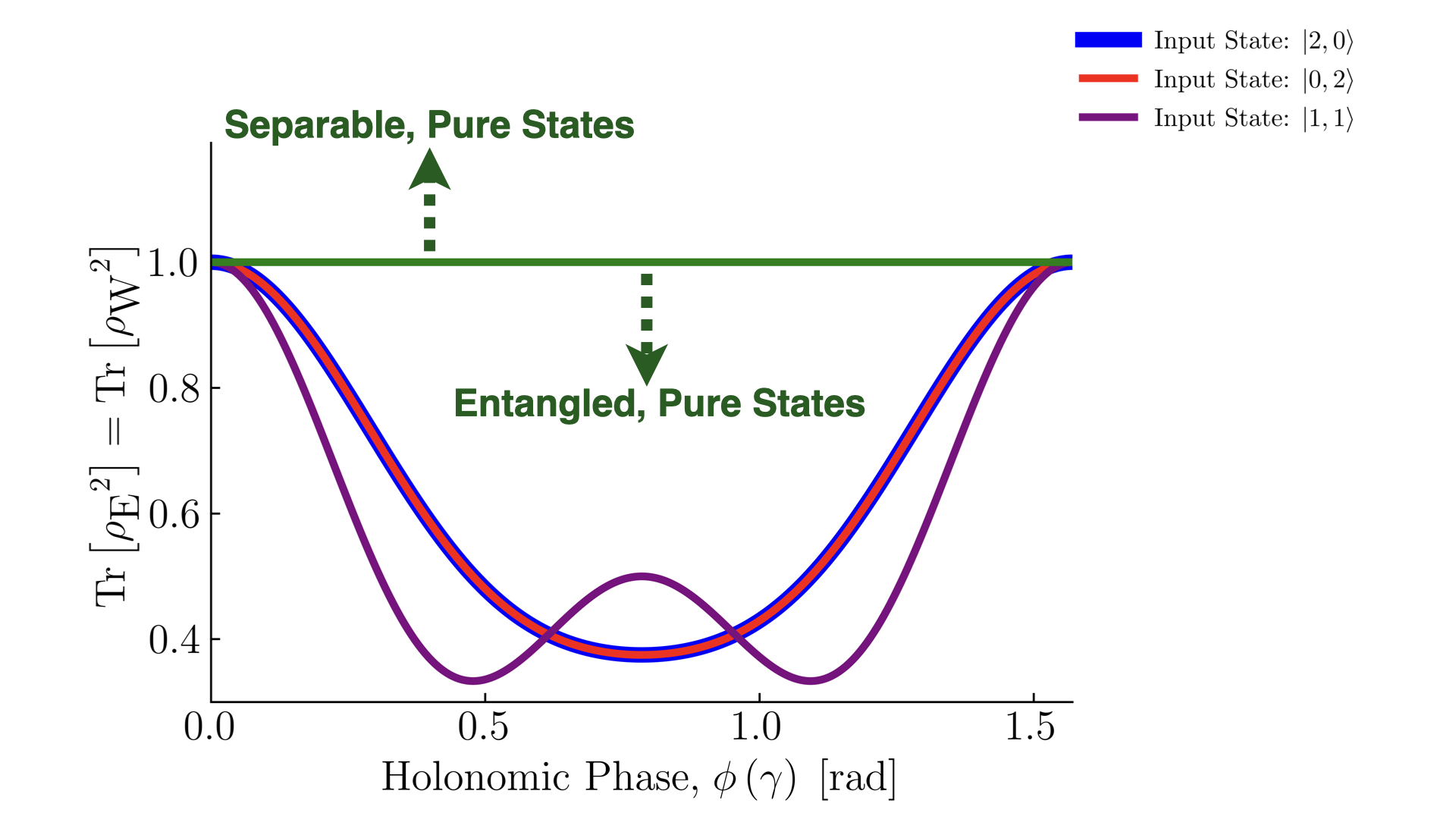}
\caption{\label{fig:fS2} \textbf{Demonstrating pure-state, bipartite entanglement using the second-order R\'enyi entropy.} The variation in the quantum purity—that is, the trace of the square of the reduced density matrix of one of the subsystems—$\textrm{Tr}\left[\rho_{\textrm{E}}^2\right]=\textrm{Tr}\left[\rho_{\textrm{W}}^2\right]$ as a function of the non-Abelian phase, $\phi\left(\gamma\right),$ which is accumulated due to adiabatic propagation of the photons through the entire holonomic chip. The second-order R\'enyi entropy is the logarithm—usually, to the base 2—of the quantum purity with an overall negative sign. The solid blue-, red-, and purple-colored curves correspond to the situations in which the product states $\left|2,0\right>,$ $\left|0,2\right>,$ and $\left|1,1\right>$ are introduced at the input. The horizontal green line demarcates the boundary between the regions of formation of separable and entangled pure states.}
\end{figure}

Motivated by such nonlinear tests, the entanglement metric known as the R\'enyi entropy—for the individual subsystems—is usually defined as:
\begin{equation}                
S_2\left(\textrm{E}\right) = -\textrm{log}_2\textrm{Tr}\left(\rho_{\textrm{E}}^2\right); \; S_2\left(\textrm{W}\right) = -\textrm{log}_2\textrm{Tr}\left(\rho_{\textrm{W}}^2\right).
\end{equation}
Figure \ref{fig:fS2} shows the generalized behavior of the R\'enyi entropy as a function of the overall non-Abelian phase, which is accumulated due to spatial propagation through the photonic chip. The key importance of this metric stems from the fact that the quantum purity—and consequently, the second-order R\'enyi entropy—can be directly measured in the laboratory \cite{islam2015measuring, brydges2019probing}.

\begin{figure}
\includegraphics[width=\columnwidth]{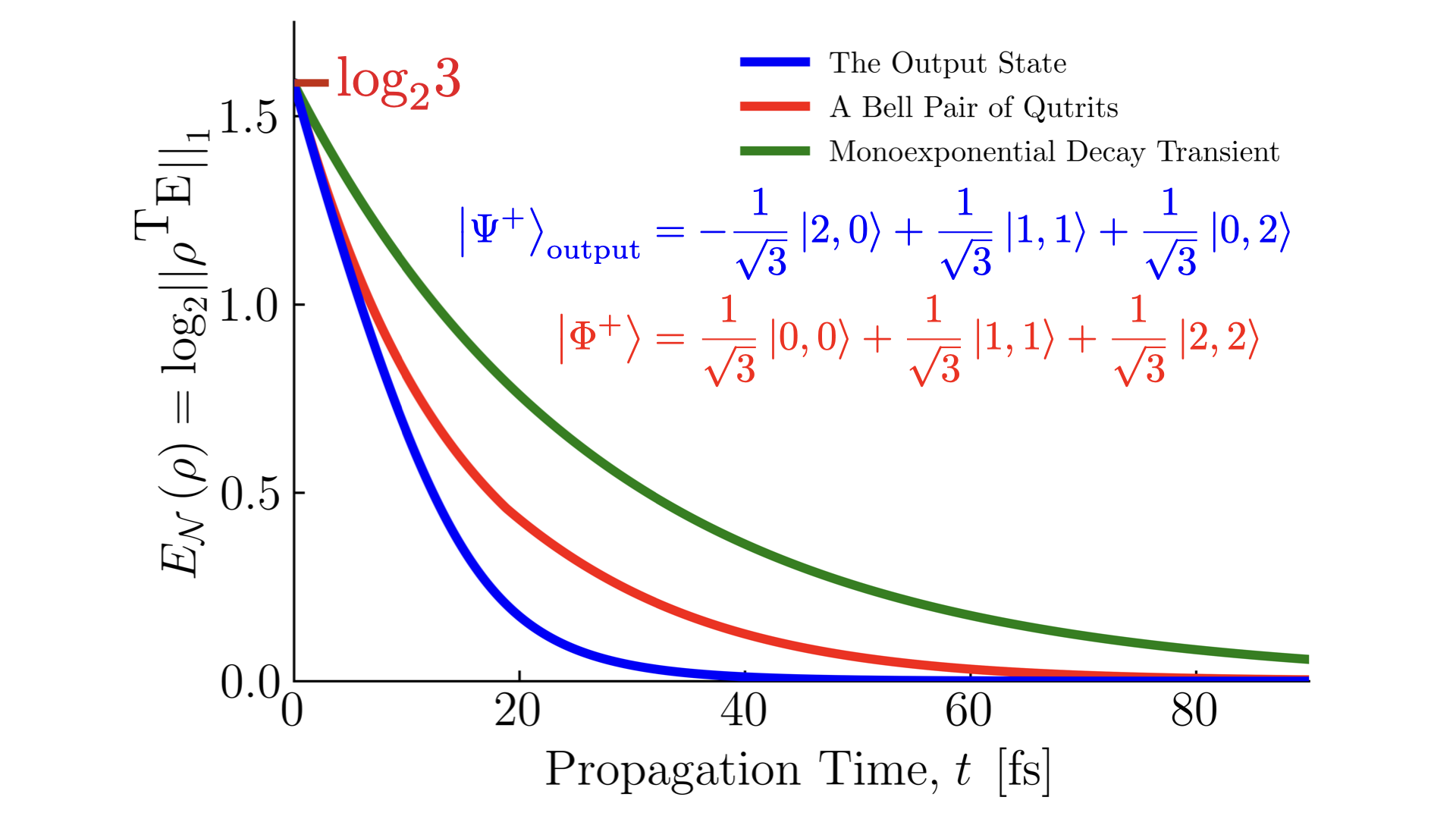}
\caption{\label{fig:fS3} \textbf{Impact of quantum erasure errors on propagating, maximally-entangled states.} The evolution of the logarithmic negativity, $E_{\mathcal{N}}\left(\rho\right) = \textrm{log}_2\vert\vert\rho^{\textrm{T}_\textrm{E}}\vert\vert_{1}$ of bipartite, initially maximally-entangled superpositions of qutrits as a function of time of propagation through a pair of waveguides, each of which is assumed to have identical single-photon dissipation rates. We chose this value of the single-photon loss rate, $\gamma = 0.1 f,$ where $f$ is the photon frequency, in order to magnify, and therefore clearly illustrate the effect of erasure errors on the entanglement dynamics. The waveguides are assumed to be continuations of the east and the west waveguides, and commence at the output of the non-Abelian holonomic device. The solid blue and red lines depict the situations in which the maximally-entangled state described by Eq. () of the main text and an equally entangled Bell pair of qutrits are introduced at the inputs of the above-described waveguide continuations, respectively. To draw a comparison between the dynamics of the mixed-state entanglement and those of single-photon dissipation, we have also plotted $e^{-\gamma t}$ as a solid green line to indicate the decay of the single-photon occupation number in each of the waveguides.}
\end{figure}

\section{Effects of Quantum Errors on Entanglement Generation} \label{sec:quantum_errors}

We consider two principal categories of quantum errors that can adversely affect our entangling protocol. First, there can be abrupt, diabatic jumps of photons—from the east and west waveguides to the central and auxiliary waveguides—during their propagation through the holonomic chip. In analogy with the Landau-Zener formula \cite{wittig2005landau}, we can estimate the single-photon diabatic error probability to be: $P_{\textrm{D}} = e^{-\sqrt{2} \Omega T} \approx 0.04,$ where $\Omega$ is the peak value of the inter-waveguide coupling coefficient—in dimensions of inverse length—and $\sqrt{2} T$ is the effective propagation length over which the jump is most probable, as $T/\sqrt{2}$ is the standard deviation of the Gaussian profiles of the inter-waveguide coupling coefficients. The above value is in reasonably good agreement—given the simplicity of our modeling—with numerically calculated values \cite{neef2023three}. Therefore, the total diabatic error probability is $\approx 8 \; \%$ for our U(3)-based entangling scheme; most likely, this error scales linearly with the size of the Hilbert space.

Second, we consider the effects of quantum erasure errors—that is, uncorrelated, single-photon loss events—outside of the holonomic chip after entangled state generation. Figure \ref{fig:fS3} shows the time-evolution of the logarithmic negativity—which constitutes an upper bound on the amount of entanglement that is distillable from bipartite mixed states \cite{PhysRevA.65.032314, PRXQuantum.2.030347}—as a maximally-entangled state described by Eq. of the main text and an equally entangled Bell pair of qutrits ($\left|\Phi^{+}\right>$) propagate through a pair of lossy waveguides. We choose $\left|\Phi^{+}\right>$ as a reference to benchmark the dynamics of the disentanglement. These waveguides are assumed to be continuations of the east and the west waveguides, and commence at the output of the non-Abelian holonomic device. The configuration having $\left|\Phi^{+}\right>$ as its initial state is more resilient to dissipation-induced-decoherence by virtue of having $\left|0,0\right>$ as one of its components.

\bibliography{Bibliography}